\documentclass[iop, numberedappendix]{emulateapj}
\usepackage{epsfig}
\usepackage{xspace}
\usepackage{amsfonts}
\usepackage{amsmath}
\usepackage{mathrsfs}
\usepackage{amssymb}
\usepackage{rotating}
\usepackage{tabularx}
\usepackage{booktabs}
\usepackage[usenames, dvipsnames]{color}

\usepackage{color}
\usepackage{ulem}
\usepackage{xspace}
\usepackage{cancel}
\definecolor{lgray}{gray}{0.85}

\newcommand{\lya} {Ly$\alpha$\xspace}

\begin{document}

\title{Modeling the L\lowercase{y}$\alpha$ Forest in Collisionless Simulations}
 
\author{Daniele Sorini\altaffilmark{1, 2},
        Jos\'e O$\tilde{\textrm{n}}$orbe\altaffilmark{1},
        Zarija Luki\'c\altaffilmark{3},
        Joseph F. Hennawi\altaffilmark{1}
        }
\altaffiltext{1}{Max-Planck-Institut f\"ur Astronomie, K\"onigstuhl 17, D-69117 Heidelberg, Germany; sorini@mpia-hd.mpg.de}
\altaffiltext{2}{Fellow of the International Max Planck Research School for Astronomy and
Cosmic Physics at the University of Heidelberg (IMPRS-HD)}
\altaffiltext{3}{Lawrence Berkeley National Laboratory, CA 94720-8139, USA}
\shorttitle{\lya Forest in Collisionless Simulations}
\shortauthors{Sorini et al.}

\begin{abstract}

Cosmological hydrodynamic simulations can accurately predict the properties
of the intergalactic medium (IGM), but only under the condition of retaining
high spatial resolution necessary to resolve density fluctuations in the IGM.
This resolution constraint prohibits simulating large volumes,
such as those probed by BOSS and future surveys, like DESI and 4MOST. 
To overcome this limitation, we present Iteratively Matched Statistics (IMS),
a novel method to accurately model the \lya forest with collisionless N-body
simulations, where the relevant density fluctuations are unresolved.  We use a small-box, high-resolution
hydrodynamic simulation to obtain the probability distribution
function (PDF) and the power spectrum of the real-space \lya forest flux.
These two statistics are iteratively mapped onto a pseudo-flux field of an N-body
simulation, which we construct from the matter density.
We demonstrate that our method can perfectly reproduce line-of-sight observables,
such as the PDF and power spectrum, and accurately reproduce the 3D flux power spectrum (5-20\%). We quantify the performance of the commonly used Gaussian smoothing technique and show that it has significantly lower accuracy
(20-80\%), especially for N-body simulations with achievable mean inter-particle separations in large volume simulations.
In addition, we show that  IMS produces reasonable and smooth spectra,
making it a powerful tool for modeling the IGM in large cosmological volumes
and for producing realistic ``mock'' skies for \lya forest surveys.

\end{abstract}


\maketitle

\section{Introduction}

The neutral hydrogen in the IGM imprints a characteristic pattern in the absorption spectra of
quasars, known as the ``Lyman-$\alpha$ Forest''. It represents an extraordinary cosmological probe, being able to trace
density fluctuations in the redshift range $0 \leq z \lesssim 6$ \citep[for a recent review, see][]{Meiksin_review}. The large number of quasars discovered
to date enables statistical analyses of the absorption spectra by considering the transmitted flux along many different lines of sight, called ``skewers''.
The measured statistical properties can be compared to theoretical models of the IGM, constraining cosmological parameters
as well as the thermal history of the IGM.
In this work, we focus on three observationally most relevant statistics of the transmitted flux:
the probability density function (PDF; \citealt{Rauch_1997})
the line-of-sight power spectrum (1DPS; \citealt{Croft_1999}),
and the 3D power spectrum (3DPS; \citealt{Slosar_2011}).
An additional motivation for studying the IGM is that it contains ~90\% of the baryons in the Universe,
acting as a gas reservoir for forming galaxies within the context of $\Lambda$CDM cosmology \citep[see, e.g.~][]{Rauch_1998}. 
Furthermore, the 3DPS can be used for an independent measurement of the Baryon Acoustic Oscillations (BAO) characteristic
scale; future increase in the number of observed quasars at redshifts $z>2$ promises tight constrains 
on the expansion history of the universe at high redshifts and other cosmological parameters \citep{Font-Ribera_2014}.

However, performing all the above mentioned studies requires not only precise observations, but also accurate theoretical modeling which
is far from being straightforward.
The \lya forest is the observational signature of neutral Hydrogen, and is set by the interplay
of gravitational collapse, expansion of the universe, and reionization processes due to the buildup of
a background of UV photons emitted by active galactic nuclei (AGN) and star forming galaxies
\citep{Cen_1994, Hernquist_1996, Zhang_1997, McDonald_2000, Meiksin_2001, Croft_2002}.
There is no analytic solution for the small-scale evolution of the (baryon) density fluctuations over time.
In order to precisely describe the behavior of the IGM, it is therefore necessary to treat the problem numerically.
In this respect, hydrodynamic cosmological simulations have led to a consistent description of the IGM  in the framework of
structure formation \citep{Cen_1994}.
However, they are computationally expensive, making it challenging to reach high resolutions.
Furthermore, available memory limits how large volume can be run in high-resolution simulations.
For example, it would be necessary to run a simulation of $\sim 1\; \mathrm{cGpc}$ on a side to probe the scales of BAO and study
their signature in the \lya forest \citep{Norman_2009, White_2010, Slosar_2009}.
The absorption lines are set by physical processes occurring around the Jeans scale, whose order of magnitude is
expected to be ~$100\;\mathrm{ckpc}$ \citep{Gnedin_1996, Gnedin_Hui_1998, Rorai_2013, Kulkarni_2015}.
Recent work indicates that a resolution of $20\,\mathrm{ckpc}$ is required to achieve $\sim$1\% precision in the
description of the statistics of the \lya forest \citep{Lukic_2015}. This implies that IGM BAO simulations would
require at least 50000$^3$ resolution elements to span such a wide dynamic range, far beyond current (and near future)
computational resources.

Collisionless simulations neglect baryonic pressure, therefore they are not as accurate as hydrodynamic simulations 
on small scales. However, on large scales baryonic forces are negligible,
thus collisionless simulations are as good as hydrodynamic ones in this regime.
For this reason, N-body collisionless simulations are often used in 
cosmology to study the formation and evolution of structure in large volumes, but with poor 
mean inter-particle spacing (often several hundreds $\rm ckpc$).
Clearly, it is desirable to find strategies that combine the volume of collisionless but retain the accuracy of high-resolution hydrodynamic simulations.
This objective has been recognized in the past, resulting in the development of various approximate methods to predict
the \lya forest from N-body simulations.

The simplest approach is assuming that baryons perfectly trace dark matter (DM; e.g. \citealt{Petitjean_1995, Croft_1998}).
In this over-simplified picture, the baryon density field is the scaled version of the matter density field.
However, DM particles are collisionless, so the pressure of baryons which competes with gravitational collapse is simply neglected.
The effect of pressure was instead included by e.g. \cite{Gnedin_Hui_1998} as a modification of the gravitational potential.
A different widely used strategy is mimicking baryon pressure by smoothing the matter density field with a Gaussian
kernel \citep{Gnedin_1996, Meiksin_2001, Viel_2002, Viel_2006, Rorai_2013}.
The flux field produced by the smoothed density field can then be computed imposing a polytropic temperature-density relation to the IGM \citep{Hui_1997}.
This method reproduces the statistics of the flux field reasonably well.
For example, \cite{Meiksin_2001} claim 10\% agreement between Gaussian-smoothed collisionless simulations and hydrodynamic
simulations in the cumulative distribution of the flux.

A more refined way to reconstruct the baryon density is  applying \textit{ad hoc} transformations
to the matter density field, calibrated with a reference hydrodynamic simulation \citep{Viel_2002}. Recently, \cite{Peirani_2014} exploited
a hydrodynamic simulation to calibrate a mapping from the density field of an N-body simulation to the \lya forest flux, tuned
to reproduce the PDF of the flux. Then, artificial flux skewers are created in order to reproduce the two-point function of the
flux given by the calibrating simulation. This is done first by computing the conditional PDF of the flux, given the DM density,
from the reference simulation. Subsequently, each pixel is assigned a value of such ``conditional flux''. This procedure seems
to yield reasonable correlation functions, but noisy skewers as well. This problem is remedied by drawing flux values from the
Gaussianized percentile distribution of the conditional flux, and then forcing it to match the PDF of the conditional flux.
Visually examining the plots of the resulting flux power spectrum, it appears close to the one provided by the reference
hydrodynamic simulation, but the accuracy is not quantified by the authors.

The lack of quantitative assessments in the literature makes it harder to compare the results obtained via different methods.
Conducting a more quantitative study is important for establishing which problems can be addressed by what methods.
Another important point regards the value of the filtering scale generally adopted in the Gaussian smoothing of matter.
The value of the filtering scale has not yet been measured from observational data \citep{Rorai_2013}.
So, in previous numerical studies it has been set to ``reasonable'' values, in any case not smaller than the mean interparticle spacing
of the simulations involved (otherwise the smoothing would have negligible effect). For example, \cite{White_2010} use
$139\,{\rm ckpc}$ as smoothing scale in a simulation with a box size of $1.02 \,{\rm cGpc}$ and 4000$^3$ particles.
Other authors have chosen larger values, for example \cite{Peirani_2014} compare their method with Gaussian-smoothed DM
simulations with a filtering scale of $300\, h^{-1} \rm ckpc$ and $1\,  h^{-1} \rm cMpc$.

In this work, we use the Gaussian smoothing technique as a starting point
upon which we add more refined transformations of the matter density field.
Following this line of reasoning, we develop two methods, named 1D-IMS and
3D-IMS, where IMS stands for ``Iteratively Matched Statistics'', the
technique on which they are grounded. The purpose of our methods is to accurately
obtain the flux statistics from collisionless simulations. This is done
through hydro-calibrated mappings, which are conceptually simpler than
those adopted by \cite{Peirani_2014}. We quantify how accurately our methods
reproduce the PDF, 1DPS and 3DPS of the flux given by a reference hydrodynamic simulation.

The high accuracy of our methods and a weak dependence on the initial smoothing scale,
represents a clear advantage over the Gaussian smoothing technique. 
Our methods thus enable using large-box collisionless simulations which do not resolve the Jeans scale. 
One important application we have in mind is modeling the BAO signature in the \lya forest.  However, there are
more topics which can benefit from it: studies of UV background fluctuations, cross-correlations between
galaxies and \lya forest and others.

The paper is organized as follows.
In \S~\ref{sec:simulations} we describe our simulations and the calculation of \lya flux.
In \S~\ref{sec:limitations} we discuss the impact of the most important assumptions underlying approximate techniques to
predict the \lya forest in collisionless simulations.
The Gaussian smoothing method is explored into great detail and we present a first quantitative analysis of its
accuracy in reproducing the 3DPS of flux, as a function of the smoothing length.
In \S~\ref{sec:IMS} we describe 3D-IMS and 1D-IMS, assessing their accuracy.
We compare the performances of the various methods in \S~\ref{sec:validation}.
In \S~\ref{sec:DM_application}, we apply 3D-IMS to a future relevant context: we compute the flux statistics through an N-body simulation,
calibrating the transformations involved in our technique with a smaller hydrodynamic simulation.
We show that the method retains accuracy, while at the same time we demonstrate that the Gaussian smoothing
technique does not yield accurate predictions when applied to simulations involving large boxes.
Finally, in \S~\ref{sec:conclusions} we compare the techniques considered by us with
previous work and present our conclusions, discussing possible future applications of our work as well.

\section{Simulations}
\label{sec:simulations}

The hydrodynamic simulations we use in this paper are carried out with Nyx code \citep{Almgren_2013, Lukic_2015},
while N-body runs are performed with Gadget code \citep{Springel_2005}.
Both codes employ leapfrog --- second order accurate method for integrating the equations of motion of particles.
Both codes also adopt the particle-mesh (PM) method with cloud-in-cell (CIC) interpolation for calculating gravitational forces.
On top of the PM calculation, Gadget adds gravitational short-range force using Barnes-Hut \citep{Barnes_1986} hierarchical tree
algorithm, therefore going to the higher resolution than our Nyx runs done on a uniform Cartesian grid.
Nyx, in addition to gravity, solves equations of gas dynamics using second-order accurate piecewise parabolic method.
To better reproduce the 3D fluid flow, a dimensionally unsplit scheme with full corner coupling is adopted \citep{Colella_1990}.
Heating and cooling are integrated using VODE \citep{Brown_1989} and are coupled to hydrodynamics through Strang splitting \citep{Strang_1968}.
All cells are assumed to be optically thin and radiative feedback is considered only through the UV background model, given by \cite{Haardt_Madau_2012}.
For cooling rates and further details on the physics in Nyx simulations, we refer the reader to \cite{Lukic_2015} paper.
The cosmological model assumed is the $\Lambda$CDM model with parameters consistent with the 7-year data release of
WMAP \citep{WMAP}: $\Omega_{\mathrm{m}}=0.275$, $\Omega_{\Lambda}=1-\Omega_{\mathrm{m}}=0.725$,
$\Omega_{\mathrm{b}}=0.046$, $h=0.702$, $\sigma_8=0.816$, $n_s=0.96$.
The simulations are initialized at $z=159$ with a grid distribution of particles and Zel'dovich approximation \citep{Zeldovich_1970}.

To recover the absorption spectra from our simulations, we choose the lines of sight, which we refer to as ``skewers'',
drawn parallel to one of the sides of the simulation box.
The optical depth is computed according to equation \eqref{eq:tau}.
After extracting skewers, we rescale the optical depth so that the mean flux is $\langle F \rangle = 0.68$ at $z=3$,
a value consistent with current observations \citep{Faucher-Giguere_2008, Becker_2013}.
Unless otherwise indicated, the results presented in this work refer to redshift $z=3$, but in order to confirm our conclusions are
not dependent on redshift, we have also analyzed redshifts $z=2$ and $z=4$.

\begin{figure*}[t!]
\begin{center}$
\begin{array}{cc}
\includegraphics[width=0.95\columnwidth]{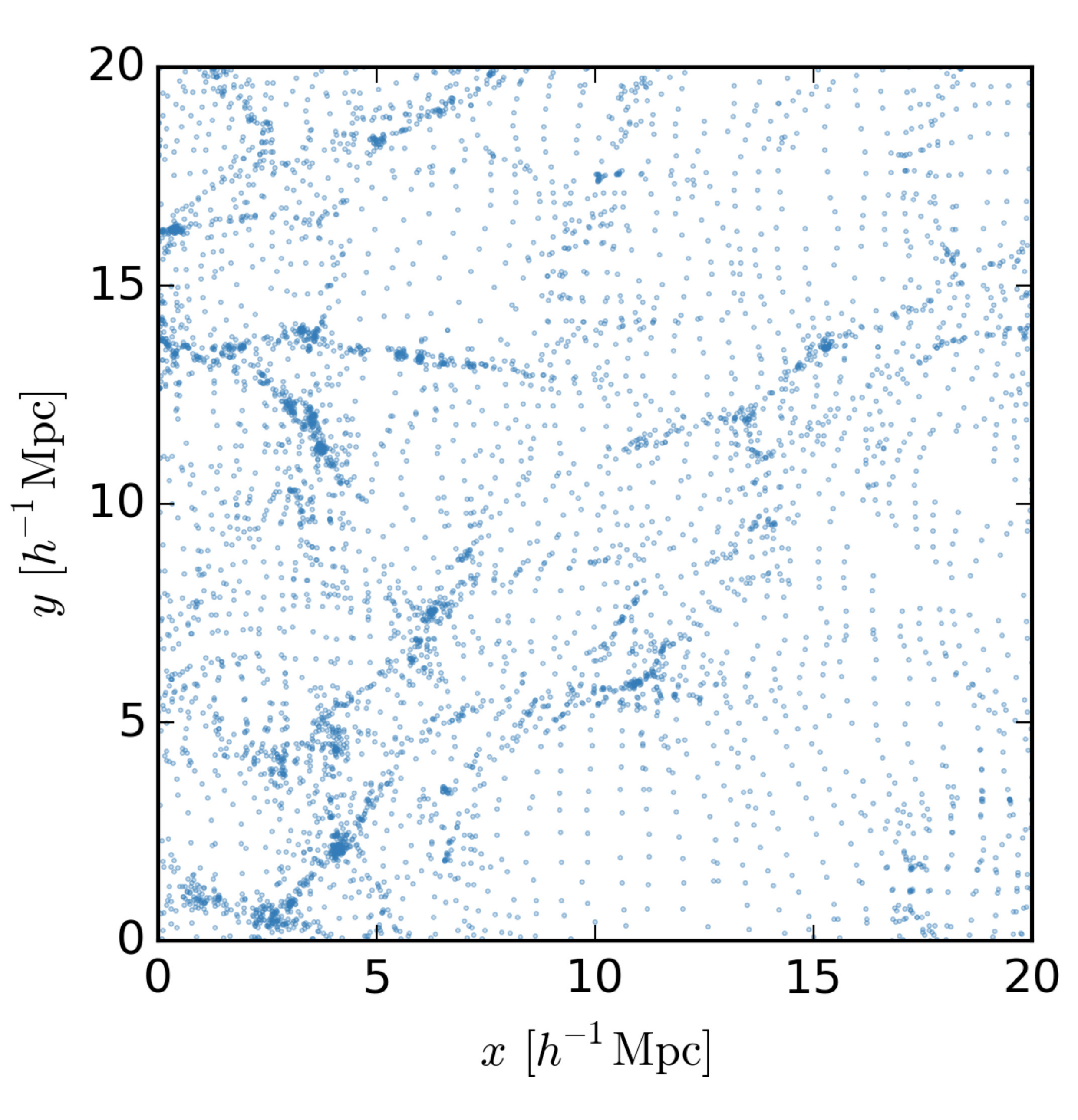} &
\hspace{0.1\columnwidth}
\includegraphics[width=0.95\columnwidth]{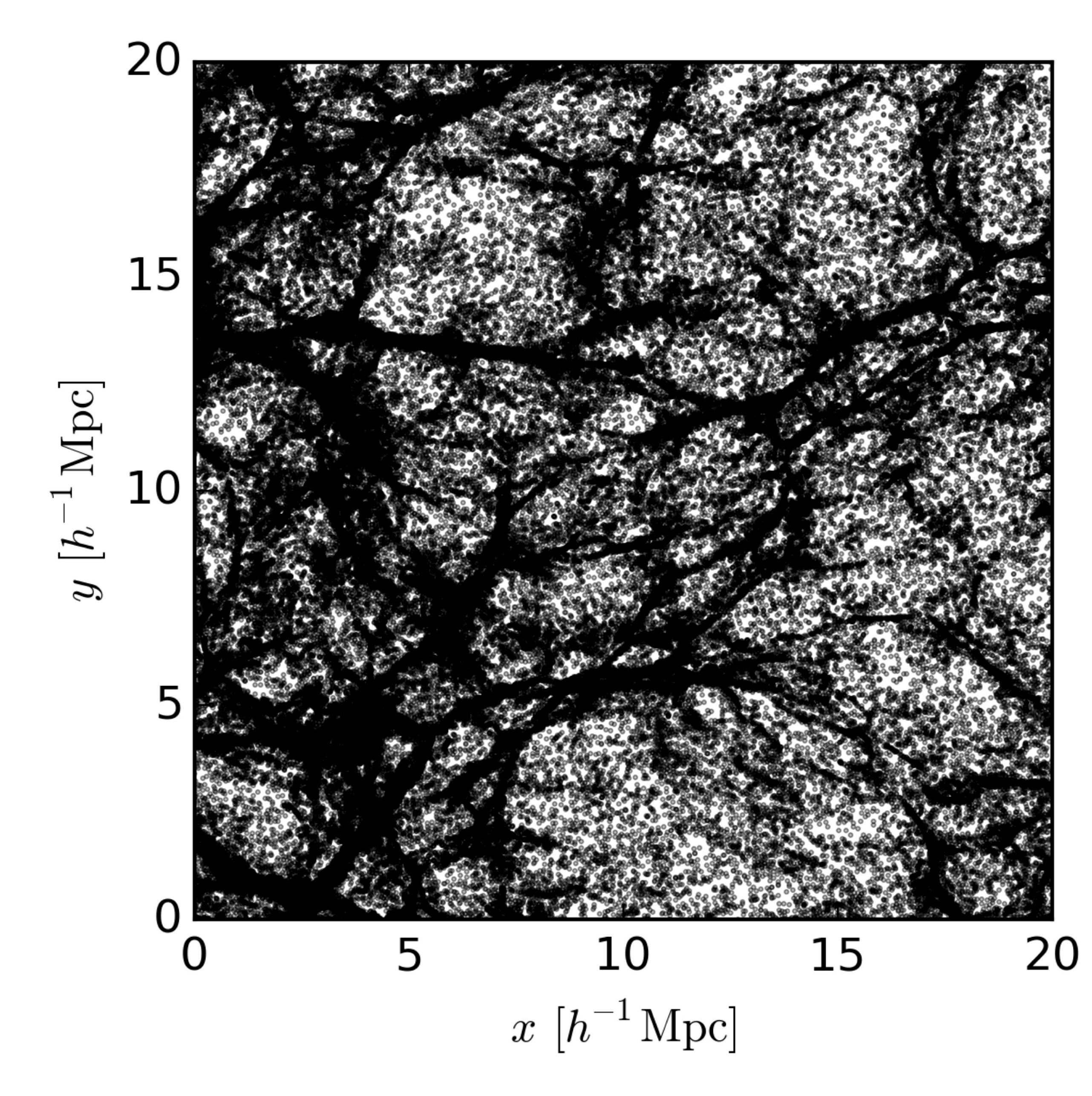}
\end{array}$
\end{center}
\vspace{-0.2in}
\caption{Difference in mass resolution between a state-of-the-art ``Hubble-volume'' simulation and a hydrodynamic simulation targeting the \lya forest. On the left (blue points) we show all particles in 0.5 $h^{-1} \rm Mpc$ thick slice from the $256^3$ Gadget N-body run. That corresponds to a trillion particles in a $4\, \rm cGpc$ box simulation. Right panel (black points) shows only 1\% of particles in the same region from the $4096^3$ Nyx hydrodynamic run, visually demonstrating the level of detail needed to capture flux statistics at percent-level accuracy.} 
\label{fig:example}
\end{figure*}

In this work we use two hydrodynamic and two N-body simulations.
The two Nyx hydrodynamic simulations have identical physics and the same spatial resolution, differing only in the choice of the box size: the smaller one has a box of $14.2\,\mathrm{cMpc}$ on a side, while the larger one $114\, \rm cMpc$.
The two simulations have $512^3$ and $4096^3$ resolution elements respectively, and they were a part of the convergence study done in \cite{Lukic_2015}.
We will first use only one Nyx simulation to test how well we can reproduce the forest statistics given only DM particles and no gas information.
For convenience, we use the smaller Nyx simulation, which is $\sim$1\% converged resolution-wise, but the box size is too small 
for accurate reproduction of flux statistics.
However, the main point of our work is to test how well we can match the given flux statistics, and it is irrelevant how accurately that statistics is
describing a particular cosmological model.
In other words, it is important to have resolution good enough to correctly capture small-scale physics,
but it does not matter that the large-scale power is missing in the simulation.

We also want to test how accurately the forest statistics can be reproduced in large-volume simulations,
i.e. with box sizes of $1\, \rm cGpc$ and larger. Of course, we do not have the ``true'' answer for such large boxes as it
would be obtained with hydrodynamic simulations.
Instead, we will use the results of the large-box Nyx run, which is demonstrated to be converged both in 
resolution and box size \citep{Lukic_2015}, as the ``truth'', and we will reconstruct its flux statistics using an
N-body Gadget run in same box together through the small-box Nyx run. We ran two Gadget N-body simulations with the
same box size as the larger Nyx simulation, $114\, \rm cMpc$, but with only $512^3$ and $256^3$ particles.
The number of particles in Gadget runs is chosen to be representative of the mean inter-particle spacing
in the state-of-the-art N-body simulations of ``Hubble'' volumes (e.g \citealt{Habib_2012, Habib_2013, Skillman_2014}).
As an example, we show in the left panel of Figure \ref{fig:example} all particles in  a $0.7 \, \rm cMpc$ thick slice from
the $256^3$ Gadget run.  This run in $114\, \rm cMpc$ box yields approximately the same mass resolution as one trillion particles in 
a $4\, \rm cGpc$ simulation would.
The right panel displays only $1\%$ of the particles in the same region, from the $4096^3$ Nyx hydrodynamic run.
The comparison of the two panels makes it immediately apparent the high resolution which is needed for modeling
the \lya forest statistics at $\sim 1\%$ accuracy.

Gadget simulations share the same phases in the initial conditions as the large Nyx run (as clearly visible in Figure \ref{fig:example}),
enabling comparison of individual skewers. We emphasize that although somewhat artificial, this test is actually more
difficult than the real-world situation, and thus we expect that we can only overestimate the error of our method.
The reason is that in reality we would use a fully converged $\sim 100\, \rm cMpc$ hydrodynamic simulation to model flux
statistics in $\sim1\, \rm cGpc$ N-body simulations, making box size errors negligible, whereas here we cannot avoid them.

\subsection{\lya Skewers}

The \lya forest arises from the scattering of photons along their path from a background quasar to the observer.
The fraction of the transmitted flux is $F=\exp(-\tau)$, where $\tau$ is the opacity of the intervening IGM.
The opacity in redshift space at a given velocity coordinate $u$ along the line of sight is given by
\begin{equation} \label{eq:tau}
	\tau (u) = \int  du' \, \frac{\lambda_{\mathrm{Ly}\alpha} \sigma \, n_{\mathrm{H I}}(\boldsymbol{u'})}{H(z) b(\boldsymbol{u'})} \exp\left[-\frac{(u-u_0(\boldsymbol{u'}))^2}{b(\boldsymbol{u'})^2}\right]
\end{equation}
where $u'$ is the component of the Hubble flow velocity field $\boldsymbol{u'}$  along the line-of-sight, over which the integral is calculated.
In the above expression, $n_{\mathrm{HI}}(\boldsymbol{u'})$ is the number density of neutral hydrogen and
$\sigma$ and $\lambda_{\mathrm{Ly}\alpha}$ are the cross section\footnote{Actually, when one considers thermal motions of the gas particles,
the cross section $\sigma_{\mathrm{Ly}\alpha}$ of the \lya transition is given by the product of $\sigma$ and a Voigt profile.
Integrating $\sigma_{\mathrm{Ly}\alpha}$ over all possible frequencies of the intervening photon, one obtains $\sigma$.
Therefore, strictly speaking, $\sigma$ is the frequency-integrated cross section and, as such, has the dimensions of area/time.
For an extensive derivation of equation \eqref{eq:tau}, see e.g. \cite{Meiksin_review}.}
and wavelength of the \lya transition in the rest frame, respectively.
The line-of-sight velocity of gas particles is then given by $u_0(\boldsymbol{u'}) = u' + u_{\mathrm{pec}}(\boldsymbol{u'})$,
the second term being the peculiar velocity of the gas. In equation \eqref{eq:tau} thermal broadening is described
by $b(\boldsymbol{u'})=\sqrt{2 k_B T(\boldsymbol{u'})/m_p}$, where $T(\boldsymbol{u'})$ is the temperature of the gas and $m_p$ the proton mass.
The convolution with thermal broadening and peculiar velocities actually yields a Voigt profile in equation \eqref{eq:tau} instead of a Gaussian.
However, the latter is a good approximation for $\tau < 100$ \citep{Lukic_2015}, regime relevant for the \lya forest studies.
Computation of $\tau$ requires a determination of the neutral hydrogen density, which in turn depends on baryon density and temperature,
as well as the hydrogen ionization and recombination rates.  The challenge for approximate methods is to recover relevant \lya forest
statistics, without the knowledge of baryon thermodynamical quantities.

\section{Limitations of Approximate Methods}
\label{sec:limitations}

The very first task of approximate methods is to obtain an estimate for the baryon density field.  This is commonly done
via manipulation of the density field in an N-body run \citep{Meiksin_2001, Viel_2002, Peirani_2014}, 
to account for the baryonic pressure smoothing. 
The functional form for the smoothing is usually a Gaussian and that is indeed the starting point for all methods considered in this paper.
Secondly, approximate methods need other assumptions, concerning the estimate of the
temperature of the IGM and its velocity field. In this section, we review these
approximations and assess their impact on the accuracy of the \lya flux, as a function of the smoothing length.
In order to better understand inherent limitations of the approximate methods, we will also consider separately the
accuracy of baryon density reconstruction of other thermodynamic quantities.

\subsection{Gaussian Smoothing}
\label{sec:Gauss_smoothing}

A pseudo baryon density field can be generated from a collisionless simulation by smoothing the matter density fluctuations
$\delta$ at a characteristic smoothing length $\lambda_G$ given as
\begin{equation}
\label{eq:gauss_smooth}
	\delta^{\lambda_G} (\boldsymbol{k}) = \delta(\boldsymbol{k}) \exp ( -\lambda_G^2 k^2 ) \, .
\end{equation}
This length is expected to be of the order of the Jeans filtering scale, which is in comoving units \citep{Binney_2008}:
\begin{equation}
	\lambda_J^2(t) = \frac{c_s^2(t)a(t)}{4\pi G \rho_{\mathrm{0}}} \, \, ,
\end{equation}
where $c_s$ is the speed of sound at time $t$, $a(t)$ the scale factor and $\rho_{\mathrm{0}}$ the mean
matter density and $G$ is Newton's gravitation constant. The same line of reasoning can be applied to
the line-of-sight velocities of particles as well. Once both matter density and velocities are smoothed, flux skewers can be
computed with some approximation for the IGM temperature (discussed in \S~\ref{sec:FGPA}),
replacing baryon density and velocity fields with the corresponding smoothed matter quantities. For the
sake of clarity, we summarize the inputs required to apply the Gaussian smoothing technique (and our methods, which will be
discussed in \S~\ref{sec:3D-IMS} and \S~\ref{sec:1D-IMS}) in Table \ref{tab:inputs}.

There are quantitative studies in the literature aiming to understand how well the Gaussian smoothing technique
reproduces various flux statistics computed through hydrodynamic simulations (see \S~\ref{sec:comparison_past} for details),
but none of them considers the flux 3DPS. We take $\lambda_G$ as a free parameter and assess the accuracy with which the
Gaussian smoothing technique recovers the flux 1DPS, PDF and 3DPS, through the following steps:
\begin{enumerate}
\item We have particle positions and velocities from a simulation.
      We deposit them on a grid using CIC deposition; we use the grid with as many cells as the number of particles in the simulation.
\item We smooth this density field with a certain smoothing scale $\lambda_G$ and the velocity field at $228\,\textrm{ckpc}$
      (see appendix \ref{app:smoothing_vel}).
\item We compute 1DPS, 3DPS and PDF of the flux field obtained using Fluctuating Gunn-Peterson Approximation (see \S~\ref{sec:FGPA}).
\end{enumerate}

\subsection{Fluctuating Gunn-Peterson Approximation}
\label{sec:FGPA}

Equation \eqref{eq:tau} can be simplified expressing the neutral hydrogen density $n_{\mathrm{H I}}$ as a function of
the baryon density fluctuations $\delta_{\mathrm{b}}$. Let us consider a gas composed by hydrogen and helium.
Let $x_{\mathrm{H II}}$, $x_{\mathrm{HeII}}$ and $x_{\rm HeIII}$ be the fractions of ionized hydrogen, singly and doubly
ionized helium respectively. The total number densities of hydrogen $n_{\rm H}$ and helium $n_{\rm He}$ are related
through $n_{\rm He}=\chi n_{\rm H}$, where $\chi=X/4Y$. Assuming photoionization equilibrium, the number density of
neutral hydrogen is given by
\begin{equation}
\label{eq:photoionization_eq}
	n_{\mathrm{H I}} = \frac{\alpha(T)}{\Gamma _{\mathrm{H I}}} x_{\mathrm{H II}} [(1+\chi) x_{\mathrm{H II}}+ \chi x_{\mathrm{He III}}] n_{\mathrm{H}}^2
\end{equation}
where $\alpha(T) \propto T^{-0.7}$ is the Case A recombination coefficient per proton and $\Gamma _{\mathrm{H I}}$ the
photoionization rate of hydrogen. Commonly used Case A and B definitions differentiate media that allow 
the Lyman photons to escape or that are opaque to these lines (except for Lyman-alpha) respectively.
Case A is more appropriate for this reionization calculation, because most of the photons produced through recombination
lie in regions of dense and partially neutral gas, so they are immediately re-absorbed and do not really contribute to the
ionizing background (\citealt{Furlanetto_2006, Kuhlen_2012}; see also the discussion in \citealt{Miralda_2003} and \citealt{Kaurov_2014}).
If helium is only singly ionized, the factor between square brackets in equation \eqref{eq:photoionization_eq} becomes
$(1+\chi) x_{\mathrm{H II}}$, while for $x_{\mathrm{H II}}=x_{\mathrm{HeIII}}$ it is $(1+2 \chi)x_{\mathrm{H II}}$.
Apart from the detailed modeling of the ionized fractions, the important point of equation \eqref{eq:photoionization_eq} in this
context is that $n_{\rm HI} \propto T^{-0.7} n_{\rm H}^2$.

Simulations show that the temperature-density relationship of the IGM is a power law over a wide range of density
and temperature \citep{Hui_1997}, so that
\begin{equation} \label{eq:temp-dens relation}
	T(\boldsymbol{u}) = T_0 (1+\delta _b (\boldsymbol{u}))^{\gamma-1}
\end{equation}
where $T_0$ and $\gamma$ are constants. From our simulation, at redshift $z=3$, we obtained $T_0=1.09 \times 10^4 \,\mathrm{K}$ and $\gamma=1.56$,
following the fitting procedure described by \cite{Lukic_2015}.
Assuming \eqref{eq:temp-dens relation}, the relationship between $n_{\rm HI}$ and $\delta_{\rm b}$ can be expressed in terms
of the parameters of our simulation as follows:
\begin{multline} \label{eq:n_HI}
	n_{\mathrm{H I}}(\boldsymbol{u}) = A \frac{8.28 \times 10 ^{-13} \, \mathrm{s}^{-1}}{\Gamma _{\mathrm{H I}}} \frac{\Omega_{\mathrm{b}}h^2}{0.0227} \left(\frac{1+z}{4} \right)^3 \\  \left(\frac{T_0}{1.09\times 10^4\,\mathrm{K}}\right)^{-0.7} [1+\delta _{\mathrm{b}} (\boldsymbol{u})]^{2-0.7(\gamma-1)}
\end{multline}
where $A$ is a proportionality constant. For our simulation, $A=3.09 \times 10^{-12}\, \rm cm^{-3}$. 
Neglecting the scatter in the temperature-density relationship of the IGM, i.e. assuming \eqref{eq:temp-dens relation} and
consequently $n_{\mathrm{HI}} \propto (1+\delta_b)^{2-0.7(\gamma-1)}$, is usually referred to as
``Fluctuating Gunn-Peterson Approximation'' (FGPA; \citealt{Weinberg_1997, Croft_1998}). Since the FGPA is useful when one
cannot or does not wish to run a hydrodynamic simulation, one also needs an approximation for $\delta_{\mathrm{b}}$ in equation \eqref{eq:n_HI}.
For this reason, in any practical situation $\delta_{\mathrm{b}}$ is replaced by the DM density fluctuations $\delta_{\rm DM}$, with
or even without Gaussian smoothing. For the sake of clarity, in the remainder of our work we shall refer solely to the operation described by  equation \eqref{eq:gauss_smooth} with ``Gaussian smoothing''. On the contrary, the Gaussian smoothing of the DM density field, combined with the FGPA to compute the \lya flux field, shall be denoted as ``Gaussian smoothing and FGPA'' (GS+FGPA).

We now define a new field, the ``flux in real space'' (or simply ``real flux'') $F_{\mathrm{real}}$ as the flux that would be
obtained neglecting thermal broadening and peculiar velocities. This is not a physical observable, but the shape of its power
spectrum is sensitive to the Jeans scale \citep{Kulkarni_2015} and it will be a useful quantity in our computations. As such,
we can define the opacity in real space
\begin{equation} \label{eq:tau_real}
	\tau _{\mathrm{real}} (\boldsymbol{u})= \frac{\lambda_{\mathrm{Ly}\alpha} \sigma}{H(z)} n_{\mathrm{HI}}(\boldsymbol{u})
\end{equation}
Within the FGPA, $\tau_{\rm real} \propto n_{\rm HI} \propto (1+\delta_b)^{2-0.7(\gamma-1)}$.
Convolving \eqref{eq:tau_real} with the gas velocities and thermal broadening, one obtains \eqref{eq:tau}. 

\begin{figure*}
\centering
\includegraphics[width=\textwidth]{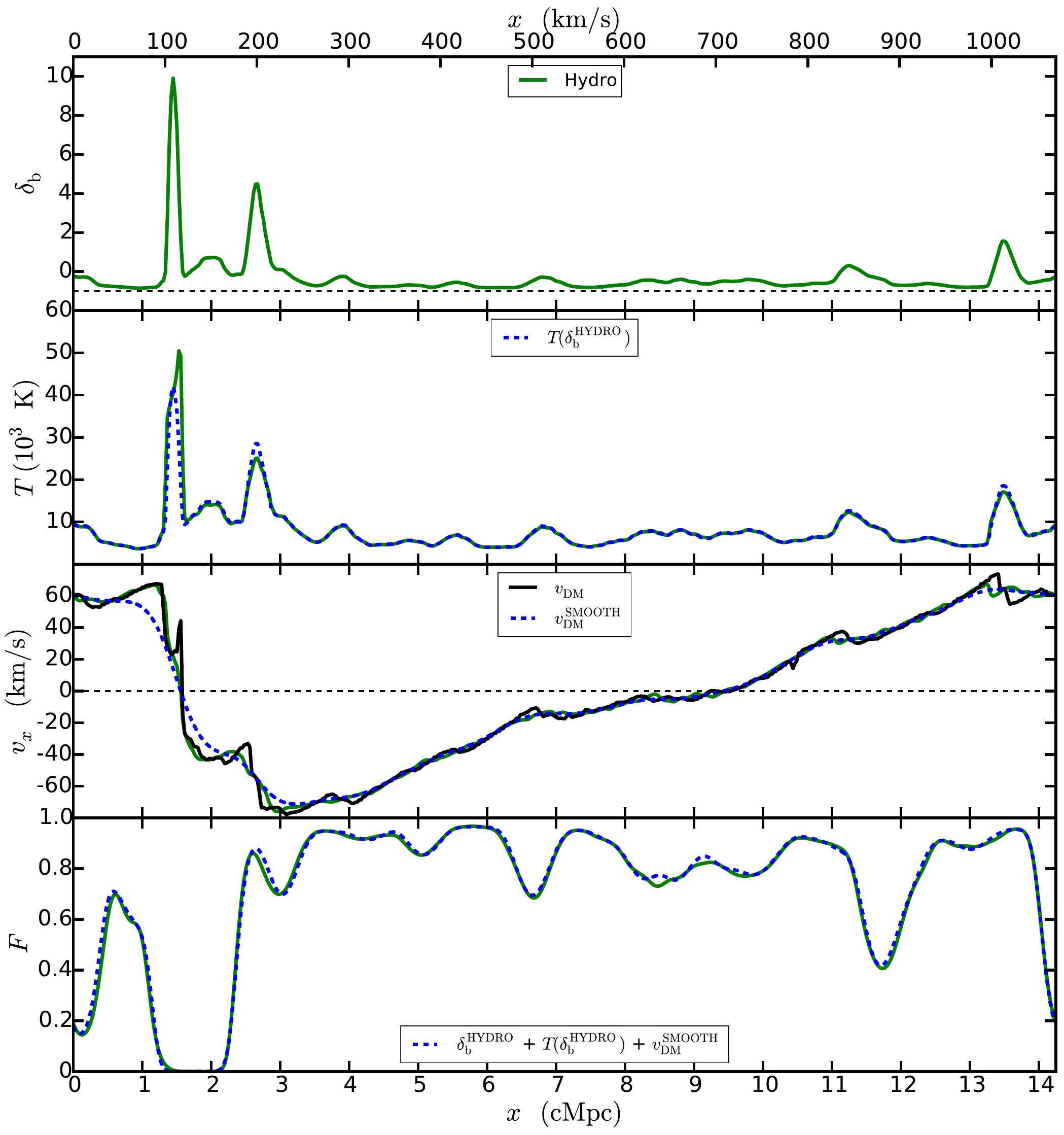}
\caption{Different quantities along a certain skewer are plotted, to illustrate possible limitations of the FGPA.
         \textit{First panel}: Baryon density fluctuations from the hydrodynamic simulation.
         \textit{Second panel}: Temperature obtained from the hydrodynamic simulation (solid green line) and by imposing a 1-to-1
         temperature-density relationship (see text for details) to the baryon density given by the simulation (dashed blue line).
         \textit{Third panel}: Line-of-sight velocity of baryons (green line) and dark matter (black line), obtained directly from
         the hydrodynamic simulation. The dashed blue line represents the line-of-sight velocity obtained smoothing the DM velocity
         with a smoothing scale of $228\,\mathrm{ckpc}$.
         \textit{Fourth panel}: Flux obtained from the hydrodynamic simulation (solid green line) and the one obtained by imposing
         a deterministic temperature-density relationship to the baryon density given by the simulation, and using the Gaussian-smoothed
         line-of-sight velocities of dark matter instead of baryons (dashed blue line).}
\label{fig:skewers_FGPA}
\end{figure*}

\begin{figure*}
\centering
\includegraphics[width=\textwidth]{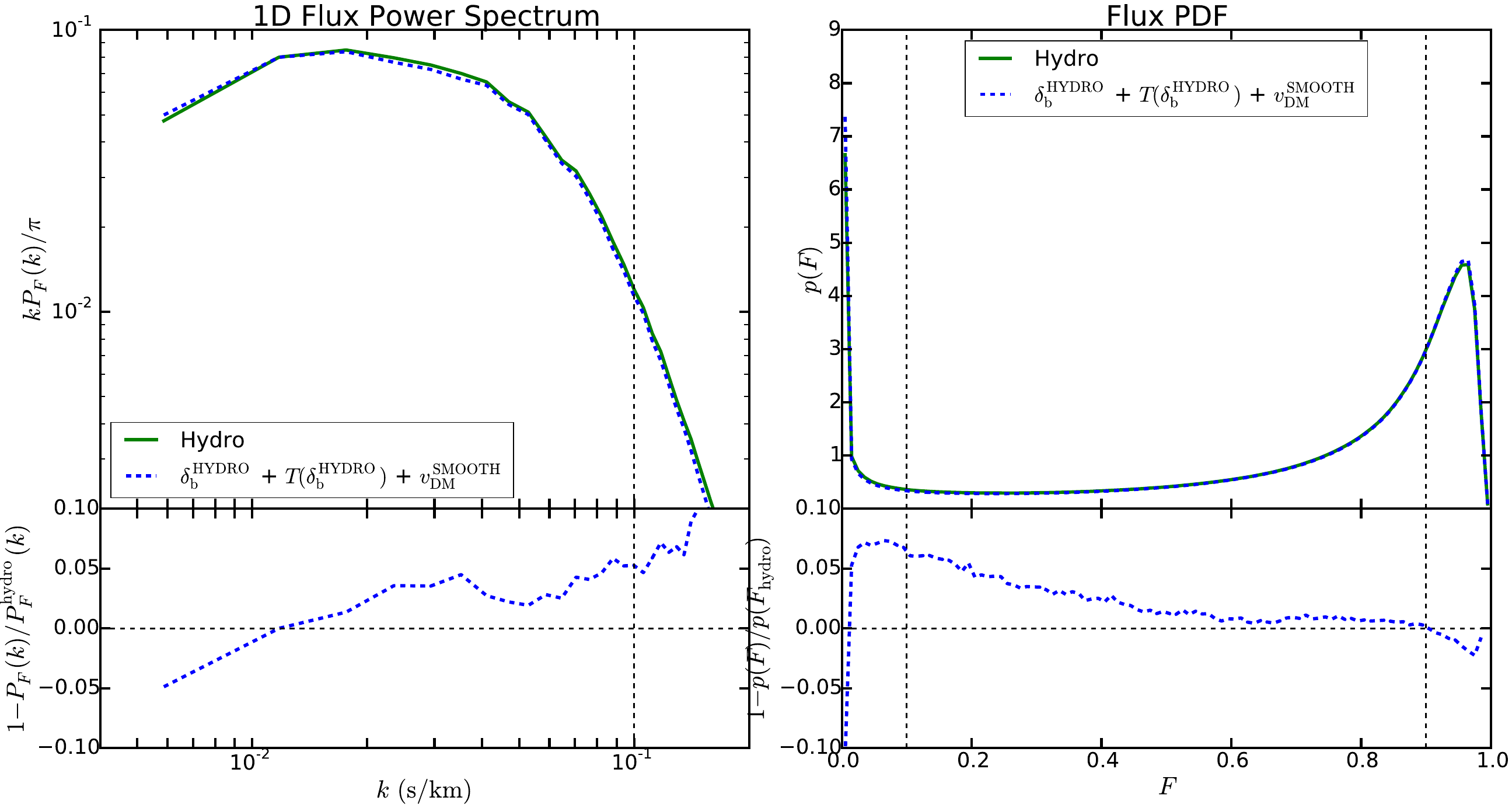}
\caption{In the top panels, the solid green lines represent the dimensionless 1DPS (left) and
         PDF (right) of the flux given by our reference hydrodynamic simulation. The dashed blue lines are the 1DPS
         and PDF of the flux computed by imposing a 1-to-1 temperature-density relationship on the baryon density given
         by the hydrodynamic simulation, and using the Gaussian-smoothed line-of-sight velocities of dark matter instead of baryons.
         The dashed vertical line delimits the dynamic range considered to compute the accuracy (see text for details).
         The relative errors plotted in the lower panels set the intrinsic limitations of approximate techniques
         predicting the \lya forest through the manipulation of the DM density field given by collisionless simulations.} 
\label{fig:FGPA_accuracy}
\end{figure*}
\begin{figure*}
\centering
\includegraphics[width=\textwidth]{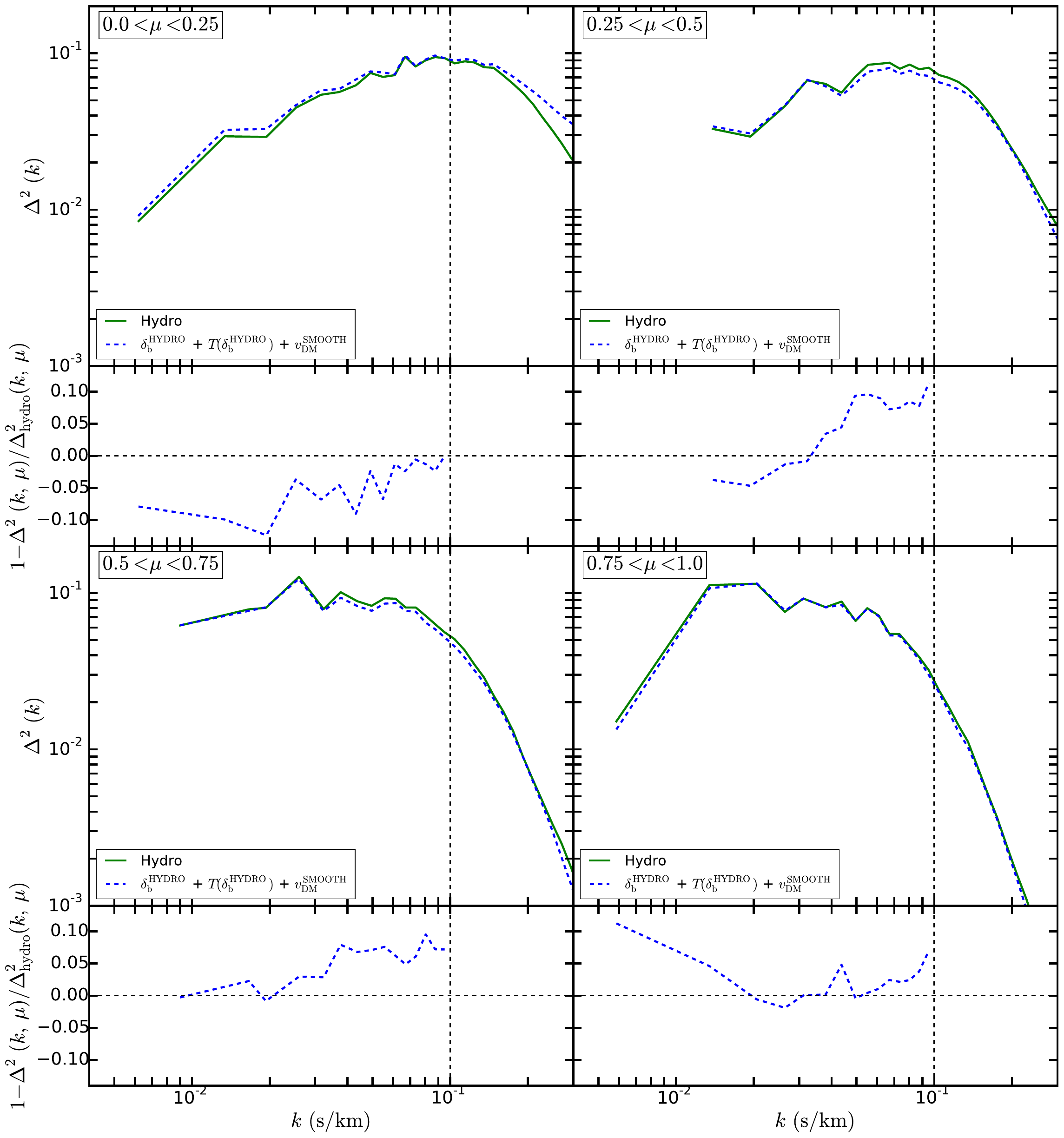}
\caption{We show the dimensionless 3DPS $\Delta^2(k,\,\mu)$ of the flux given by our reference hydrodynamic
         simulation (solid green lines) and of the flux computed by imposing a 1-to-1 temperature-density relationship on
         the baryon density given by the hydrodynamic simulation, and using the Gaussian-smoothed line-of-sight velocities
         of dark matter instead of baryons (dashed blue lines). We consider 4 bins of $\mu$, and show 
         $\Delta^2(k,\,\mu)$ as well as the relative difference between the spectra. The dashed vertical line marks the dynamic
         range considered to compute the accuracy of the FGPA (see text for details). The relative errors plotted show the
         intrinsic limitations of approximate techniques predicting the \lya forest through the manipulation of the
         DM density field given by collisionless simulations.} 
\label{fig:3DPS_FGPA}
\end{figure*}

\subsection{Accuracy of FGPA}
\label{sec:FGPA_accuracy}

We now assess the accuracy of the FGPA using the baryon density field from the hydrodynamic simulation as a reference.
The motivation is that approximate methods are based on smoothing both DM density and sometimes also velocity field at a
certain scale set by a Gaussian kernel. This is meant to mock up the smoothing
of baryons due to their finite pressure \citep{Gnedin_Hui_1998, Kulkarni_2015}.
But we are interested in ``inherent'' accuracy of FGPA, thus we will use 
the actual baryon density field from the simulation, and focus on the effects of the velocity smoothing
and the assumption that the temperature-density relationship is a power law.

In our hydro simulation, we also have the velocities of DM particles.
Thus we construct the velocity field by CIC-binning them on a grid with as many cells as the number of particles and then smooth it
with a Gaussian kernel. In principle, the smoothing length of velocity could be different from the one of the DM density field.
We keep it fixed to $228\,\mathrm{ckpc}$ throughout the paper, since we verified that this value gives the best overall accuracy
in reproducing the statistics considered (see appendix \ref{app:smoothing_vel} for further details).
However, we have also checked that modifying the smoothing length for the velocity field does not significantly change our conclusions.

In Figure \ref{fig:skewers_FGPA} we show different physical quantities along one skewer as an example, to display the
differences between the hydrodynamic simulation (solid green lines) and the FGPA (dashed blue lines). The top panel shows the
density fluctuations along the skewer considered. The second panel underscores the differences between a temperature-density
relationship with no scatter and the temperature given by the hydrodynamic simulation. We see that the biggest differences arise
around the highest density peaks, where shocks could be present.
In the third panel we plot the line-of-sight velocity of DM particles\footnote{The CIC-binned velocity field of DM particles occasionally results in pixels with no particles in them. To correct for this effect, we assign to these grid cells the average velocity of their first neighbors in the 3D space. Then, we proceed with the Gaussian smoothing.} (black line) and baryons (green line). Here we also plot the smoothed DM velocity, which is the one we actually adopt (dashed blue line).
In the last panel, we show the difference between the flux computed as explained in this section and from the hydrodynamic simulation.
We notice that the FGPA recovers the flux skewer remarkably well.

We show the results about the statistics of flux skewers in Figures \ref{fig:FGPA_accuracy} and \ref{fig:3DPS_FGPA}.
In the upper panels of Figure \ref{fig:FGPA_accuracy} we show the flux 1DPS and PDF given by the hydrodynamic simulation and
the FGPA applied as explained above. In the lower panels, we show the relative difference of the statistics obtained with
respect to the results of the reference simulation. Analogous plots for the flux 3DPS can be seen in Figure \ref{fig:3DPS_FGPA}.
We recall that the 3DPS can be expressed as a function of the norm of the $\boldsymbol{k}$-mode considered and of
$\mu= \hat{\boldsymbol{n}} \cdot \boldsymbol{k}/k$, where $\hat{\boldsymbol{n}}$ is the unit vector parallel to the line-of-sight.
We shall denote the dimensionless 3DPS as $\Delta^2(k,\,\mu)=k^3 P_F(k,\,\mu)/2 \pi ^2$.

The accuracy of the FGPA of course depends on the Fourier modes considered for the power spectra and on the specific binning
adopted for the flux PDF. We now wish to define a set of parameters describing the overall goodness of the method.
For this purpose, we first of all delimit a range of Fourier modes and flux in which it is sensible to compare the statistics
obtained via the simulation and the FGPA. Since small scales are often contaminated by metal lines, we consider modes below
$k=0.1\,{\rm s\,km^{-1}}$ \citep{Lidz_2010}. This upper bound is indicated with the vertical dashed line in the left panels
of Figure \ref{fig:FGPA_accuracy}. The overall accuracy of the FGPA is assessed by the arithmetic mean of the modulus of the
relative error in the range of $k$ considered: 
\begin{equation}
	m  = \frac{1}{N}\sum _{k < 0.1\;\rm s\,km^{-1}} \frac{\left\vert P^{\mathrm{hydro}}_F(k) - P^{\mathrm{FGPA}}_F(k) \right\vert}{P^{\mathrm{hydro}}_F(k)}
\end{equation}
where $N$ is the number of modes in such range. A small value of $m$ implies a good mean accuracy.
Note however that it does not necessarily mean that the accuracy is good \textit{everywhere}.
Indeed, a low value of $m$ can be achieved by a set of points where the relative error is extremely close to zero for
many of them but large for just a couple of modes. In other words, $m$ tells us nothing about the \textit{dispersion}
of the relative error around its mean value. To estimate such dispersion, we simply compute the root-mean-square $s$ of
the relative error in the range considered:
\begin{equation}
	s^2 = \frac{1}{N}\sum _{k < 0.1\;\rm s\,km^{-1}} \left(\frac{\left\vert P^{\mathrm{hydro}}_F(k) - P^{\mathrm{FGPA}}_F(k)\right\vert}{P^{\mathrm{hydro}}_F(k)} - m \right)^2
\end{equation}
The range within which we compute $m$ and $s$ is $0.1<F<0.9$. The upper bound means that we are excluding a range of flux
often limited by continuum placement uncertainties \citep{Lee_2012}, whereas the lower bound translates into ignoring flux
values susceptible to inaccuracies in modeling optically thick absorbers \citep{Lee_2015}.
The same analysis is applied to the 3DPS as well, by doing a separate calculation for each bin of $\mu$
(we consider 4 $\mu$-bins, evenly spaced between 0 and 1).

The mean accuracy of FGPA at a smoothing length of $228\,\mathrm{ckpc}$ in reproducing 1DPS and PDF of the flux given by
the hydrodynamic simulation is 2\%. For the 3DPS, it is between 3\% and 5\%, depending on the $\mu$-bin considered.
We stress that these levels of accuracy are obtained employing in the computations the baryon density provided by the hydrodynamic simulation.
It means that, regardless how well we create the pseudo density field, this sets our limiting accuracy.
To improve it even more, one should come up with more refined ways of reproducing the velocity field and the scatter in the
temperature-density relationship. 

To sum up, one source of error is considering DM velocities instead of baryonic ones.
This is minimized because we looked for the optimal smoothing length for the velocity field.
The remaining uncertainty arises from the scatter in the temperature-density relationship which is not captured by the FGPA.

\section{Iteratively Matched Statistics}
\label{sec:IMS}

To better model \lya forest in collisionless simulations,
we developed two novel methods which iteratively match certain \lya forest flux statistics given as input.
The most accurate inputs today come from hydrodynamic simulations, and that is what we use here.
We name this technique ``Iteratively Matched statistics'' (IMS); the two methods are called 3D-IMS and 1D-IMS.
\begin{table*}
\caption{Inputs needed for the different methods considered.}
\label{tab:inputs}
\begin{tabular}{lcccc}
\hline
Method & DM Particle Distribution & $(\lambda_G,\,T_0,\,\gamma)$ & $F_{\rm real}$: 3D Power Spectrum and PDF & $F$: 1D Power Spectrum and PDF\\
\hline
GS+FGPA & $\checkmark$ & $\checkmark$ & & \\
3D-IMS & $\checkmark$ & $\checkmark$ & $\checkmark$ & \\
1D-IMS & $\checkmark$ & $\checkmark$ & $\checkmark$ & $\checkmark$ \\
\hline
\end{tabular}
\end{table*}

\subsection{3D Iteratively Matched Statistics}
\label{sec:3D-IMS}

The basic idea of 3D-IMS is to compute the flux from a collisionless simulation and match its one- and two-point statistics to a reference hydrodynamic simulation. Because redshift space distortions and thermal broadening make the flux field anisotropic, we for simplicity conduct this matching in real space, where the flux is an isotropic random field. So, in general, one needs a collisionless simulation and a model for the 3D power spectrum and probability distribution function of the flux in real space to apply 3D-IMS. In our case, the model for these statistics is the result of our hydrodynamic simulation. The tabulated 3D power spectrum and PDF of the  flux in real space are the inputs of the method, together with the DM particle distribution given by the collisionless simulation and the thermal parameters of the IGM (see Table \ref{tab:inputs}). 
Before going into the details of the procedure, it is worth enumerating the main steps, to better understand the logical flow.
\begin{enumerate}
\item As a starting point, the DM density is smoothed
  with a Gaussian kernel with a smoothing length $\lambda_G$, which is taken as a free parameter. In a situation where the DM was simulated on a coarse grid (e.g. with a PM code), $\lambda_G$ would be at least as large as the inter-particle simulation. The smoothed field is used to compute the flux in real space within the FGPA, following equations \eqref{eq:n_HI} and \eqref{eq:tau_real}. We shall call this flux field $F^{\mathrm{DM}}_{\rm real}$.

\item The input real flux dimensionless 3D power spectrum and PDF, taken from the hydrodynamic simulation, are used to calibrate two transformations.

  Such transformations are iteratively applied to $F^{\mathrm{DM}}_{\rm real}$, forcing its dimensionless 3D power spectrum and PDF to match the ones given as input. 
  The iterations are implemented until both statistics are matched with high precision.
\item From the resulting pseudo real flux field, a pseudo baryon density field is obtained inverting equations \eqref{eq:tau_real} and \eqref{eq:n_HI}.

\item The pseudo baryon density is Gaussian-smoothed with a smoothing length equal to the size of a grid cell. As we shall explain later, this step is necessary to remove hot pixels that give rise to non physical density skewers.  
The smoothed baryon density field is then used to compute flux skewers within the FGPA.
\end{enumerate}

\begin{figure}
\centering
\includegraphics[width=0.48\textwidth]{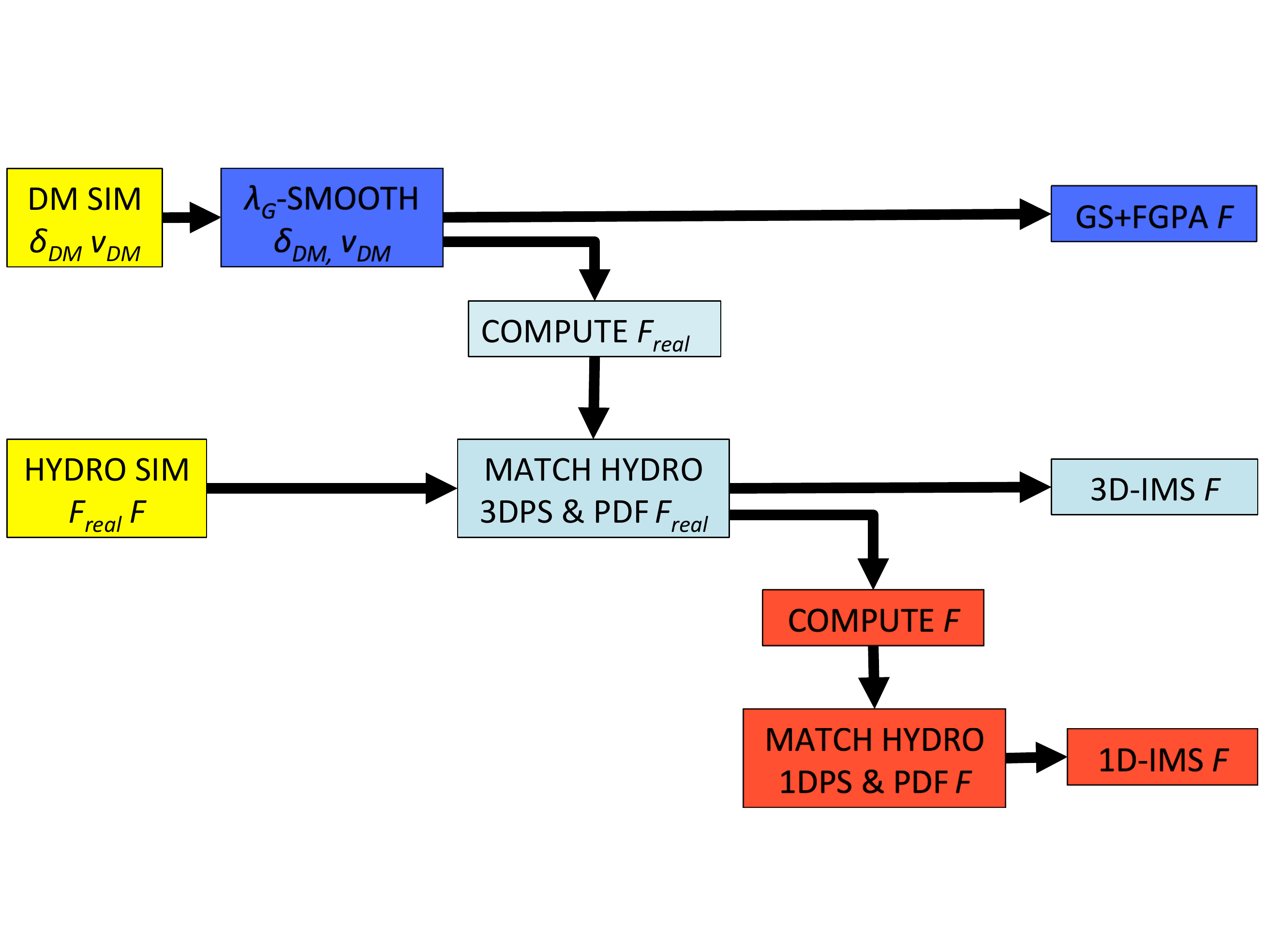}
\caption{Flow chart of the methods tested. Yellow boxes are the inputs needed. Blue boxes illustrate the steps of the FGPA applied to the Gaussian-smoothed DM density (GS+FGPA; see \S~\ref{sec:Gauss_smoothing} and \S~\ref{sec:FGPA} for details). 3D Iteratively Matched Statistics consists in appending two further steps at the end of GS+FGPA, before computing the flux field. These steps are represented by the cyan boxes. 1D Iteratively Matched Statistics requires to apply two further steps (red boxes) on top of 3D-IMS, just before extracting flux skewers.}
\label{fig:flow_chart}
\end{figure}

The points just enumerated, which can be visualized as a flow chart in Figure \ref{fig:flow_chart}, give our method its name: Iteratively Matched Statistics (IMS). The prefix 3D stresses that we are matching the dimensionless 3D power spectrum of the flux in real space. Matching this statistics is straightforward, as the $F_{\mathrm{real}}$ 3D power spectrum obeys a simple functional form \citep{Kulkarni_2015} and is isotropic in redshift space. 

On the contrary, reproducing the 3DPS of flux in redshift space would be more complicated, because it is an anisotropic power spectrum. It would require performing transformations in real space after deconvolving redshift space distortions and thermal broadening. Matching the  3D power spectrum of the baryon density would not be optimal either, since it does not exhibit an obvious Jeans cutoff \citep{Kulkarni_2015}, being dominated by higher density structures in collapsed halos at small scales. Although these rare dense regions dominate the baryon power spectrum, they contribute negligibly to variations in the \lya forest flux because the exponentiation of the opacity field maps them to zero.  As such, we choose to match the statistics of the real-space flux field, since this is an isotropic field, which is directly related to the observable, that is the flux in redshift space. 

We shall now examine the details of each step of the method.
We want to remap $F ^{\mathrm{DM}}_{\rm real}$ to a new field $F _{\mathrm{real}} ^{\mathrm{3D-IMS}}$ with the same dimensionless 3D power spectrum as $F _{\mathrm{real}}^{\rm HYDRO}$. To do this, let us consider the real flux fluctuations $\delta_{F ^{\mathrm{DM}}_{\rm real}}$ and $\delta _{F _{\mathrm{real}}^{\rm HYDRO}}$ in Fourier space. We define $F _{\mathrm{real}} ^{\mathrm{3D-IMS}}$ as
$\delta _{F _{\mathrm{real}} ^{\mathrm{3D-IMS}}} (\boldsymbol{k}) = T(k) \delta_{F ^{\mathrm{DM}}_{\rm real}} (\boldsymbol{k})$, where $T(k)$ is a function tuned to match the dimensionless 3D power spectrum of $F _{\mathrm{real}}^{\rm HYDRO}$. We shall call it ``transfer function'' and its explicit expression is given by

\begin{equation}
\label{eq:transfer}
	T(k) = \sqrt{\frac{\Delta ^2_{F _{\mathrm{real}}^{\rm HYDRO}}(k)}{\Delta ^2_{{F^{\mathrm{DM}}_{\rm real}}}(k)}}
\end{equation}
where $\Delta _X ^2(k)=k^3P_X(k)/2\pi^2$ denotes the dimensionless 3D power spectrum of field $X$. Let us point out that in our case it is straightforward to apply equation \eqref{eq:transfer}, because both $F _{\mathrm{real}}^{\rm HYDRO}$ and $F^{\mathrm{DM}}_{\rm real}$ sample the same modes, having been built from the same simulation. However, one can apply it also to the more interesting case where $F _{\mathrm{real}}^{\rm HYDRO}$ is computed from a small-box hydrodynamic simulation and $F^{\mathrm{DM}}_{\rm real}$ from a large-box N-body simulation. 
This will be discussed into more detail in \S~\ref{sec:DM_application}.

At this point, we 
compute the pseudo real flux field simply as $F _{\mathrm{real}} ^{\mathrm{3D-IMS}} (\boldsymbol{x})= \bar{F}^{\rm HYDRO}_{\mathrm{real}}(1+\delta _{F_{\mathrm{real}}^{\rm 3D-IMS}} (\boldsymbol{x}))$, where $\bar{F}_{\mathrm{real}}^{\rm HYDRO}$ is the mean value of the real flux field obtained from the hydrodynamic simulation.

The field $F _{\mathrm{real}} ^{\mathrm{3D-IMS}}$ does not have the same PDF as $F ^{\rm HYDRO} _{\mathrm{real}}$. To match the PDF, we use the argument explained by Peirani et al. (2014). We compute the cumulative distribution of both fields and we construct a mapping between the two fluxes by assigning to each value of $F _{\mathrm{real}} ^{\mathrm{3D-IMS}}$ the value of $F _{\mathrm{real}}^{\rm HYDRO}$ corresponding to the same percentile in their respective cumulative distributions. 
We now have a new pseudo real flux field $F_{\mathrm{1,}\,\mathrm{real}}^{\mathrm{3D-IMS}} (\boldsymbol{x})$, whose PDF matches by construction the one of $F _{\mathrm{real}}^{\rm HYDRO}$. However, its dimensionless 3D power spectrum is no longer the same as $\Delta ^2 _{F_{\mathrm{real}}^{\rm HYDRO}}(k)$.

To match both dimensionless 3D power spectrum and PDF of $F _{\mathrm{real}}^{\rm HYDRO}$, we iterate the two transformations. We verified that both 3D power spectrum and PDF converge to their counterparts in the simulation. This is a non trivial result.\footnote{While seeking the optimal way to match the statistics of the flux fields given by the hydro simulation, we have applied the IMS technique involving also other fields, like $n_{\rm HI}$. Convergence has not occurred in all cases.} Convergence occurs between 10 and 20 iterations, after which the improvement in the transfer function at every additional iteration is less than 0.3\%. It is worth pointing out that every time we match the 3D power spectrum there is no warranty that the new flux field has physically meaningful values, i.e. between 0 and 1. This is indeed the case, so we cannot simply compute $\delta _{\mathrm{b}} ^{\mathrm{3D-IMS}}$ from the resulting flux field. This issue is fixed naturally when we match the PDF. Since the distributions are mapped percentile to percentile and $F _{\mathrm{real}}^{\rm HYDRO}$ contains obviously only physical values, pixels with negative flux are mapped to small but positive values and pixels with flux larger than one are mapped to values close to but less than 1. It is then fundamental to conclude the iteration process matching the PDF. 

At the end of the last iteration, we have the final pseudo real flux field, whose PDF matches by construction the one of $F _{\mathrm{real}}^{\rm HYDRO}$. Since in our model there is a 1-to-1 correspondence between $\delta _b$, $n _{\mathrm{H I}}$ and $F _{\mathrm{real}}$, the PDF of the pseudo baryon density and the hydrogen number density have also converged to an asymptotic distribution. However, in the hydrodynamic simulation there is not such a correspondence, since skewers are not computed within the FGPA. As a result, the PDF of the final $\delta _{\mathrm{b}} ^{\mathrm{3D-IMS}}$ does not perfectly match the corresponding field $\delta _{\mathrm{b}}^{\rm HYDRO}$ from the reference hydrodynamic simulation. Furthermore, pseudo baryon density skewers present some non physical cuspy overdensities. They arise because the transformation matching the dimensionless 3D power spectrum of real flux introduces flux fluctuations of $\sim 10^{-3}$ in the rank ordering of pixels, which translate into large discontinuities in density in low-flux regions because of the exponentiation of equation \eqref{eq:tau_real}.
These cusps can be eliminated with a Gaussian smoothing. In this way, the density values in neighboring pixels are ``blended'' together and, as a result, very high values are turned into physical ones. The drawback is that, if we compute the real flux from the smoothed field, it will not have the same PDF as $F _{\mathrm{real}}^{\rm HYDRO}$ anymore. A good compromise is adopting the shortest possible length scale for the smoothing, that is the size of one cell of the grid on which we CIC-binned the DM particle distribution. In our case, that corresponds to $28\,\rm ckpc$. We emphasize here that this last smoothing must always be below the smallest relevant physical scale in the hydrodynamic simulation, which in our context is the Jeans scale, not to considerably affect the resulting statistics. 

Running the method for different values of $\lambda_G$, we investigate if there is a trend of the accuracy of the various flux statistics. We remind the reader that the initial smoothing serves only as a starting point for the method. In any realistic situation, the value of $\lambda_G$ is going to be related to the inter-particle separation of the underlying simulation. Indeed, smoothing on a scale smaller than that would make the PDF of the baryon density inaccurate, especially in voids \citep{Rorai_2013}, which are the most relevant regions as far as the \lya forest signal is concerned. The results of our analysis are discussed in \S~\ref{sec:validation}.

\subsection{1D Iteratively Matched Statistics}
\label{sec:1D-IMS}

The method called 1D-IMS has 3D-IMS as a starting point, on top of which further transformations are applied.  Alongside the inputs required by 3D-IMS, one needs to provide a model for the line-of-sight power spectrum 
and PDF of the flux in redhsift space as well (see Table \ref{tab:inputs}). Once again, we computed these inputs from the hydrodynamic simulation. 
After running 3D-IMS, we are left with a real flux field whose dimensionless 3D power spectrum and PDF match the ones of the real flux from the reference hydrodynamic simulation. We then compute the flux in redshift space and apply again the Iteratively Matched Statistics procedure, this time aiming at matching the dimensionless line-of-sight power spectrum and PDF of the flux in redshift space from the hydrodynamic simulation. 
As in 3D-IMS, we apply two \textit{ad hoc} transformations. Analogously to equation \eqref{eq:transfer}, we define a transfer function as follows
\begin{equation}
	T(k) = \sqrt{\frac{\Delta ^2_{F ^{\mathrm{HYDRO}}}(k)}{\Delta ^2_{{F^{\mathrm{3D-IMS}}}}(k)}}
\end{equation}
where $\Delta^2_{F ^{\mathrm{HYDRO}}}(k)$ and $\Delta ^2_{F^{\mathrm{3D-IMS}}}(k)$ are the dimensionless line-of-sight power spectra of the flux in redshift space given by the hydrodynamic simulation and obtained after running 3D-IMS respectively. After multiplying the Fourier modes of the fluctuations of $F^{\mathrm{3D-IMS}}$ by $T(k)$, we have a flux field whose dimensionless power spectrum matches $\Delta^2_{F ^{\mathrm{HYDRO}}}(k)$ by construction. 

At this point, we match its PDF to the one given by the hydrodynamic simulation exploiting the cumulative distributions, just like in \S~\ref{sec:3D-IMS}. We then reiterate the two transformations until we achieve convergence in both 1DPS and PDF. Since these statistics are now matched by construction, it would be interesting to check if the 3D correlations are preserved. We then investigate the trend of the accuracy of the 3DPS as a function of $\lambda_G$.

\begin{figure*}
\centering
\includegraphics[width=\textwidth]{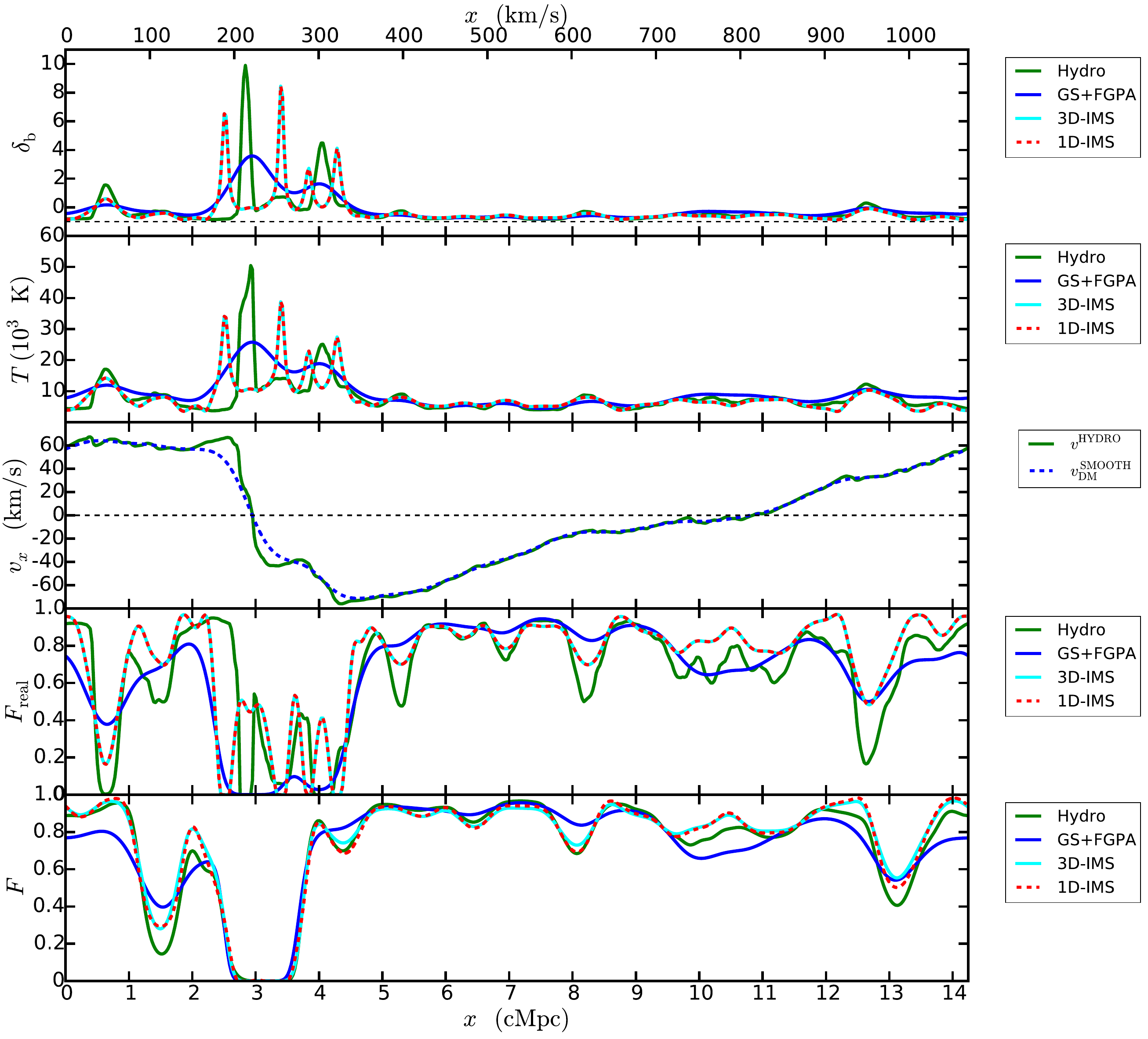}
\caption{From top to bottom, baryon density fluctuations, temperature, velocity field, flux in real space and flux in redshift space along a certain skewer. Solid green lines show results from the reference hydrodynamic simulation, solid blue lines refer to GS+FGPA, solid cyan and dashed red lines to 3D-IMS and 1D-IMS respectively. 1D-IMS consists in matching the dimensionless line-of-sight and the PDF of the flux in redshift space on top of the results given by 3D-IMS, so these two methods differ only for this quantity. The dashed blue line in the third panel from top represents the Gaussian-smoothed line-of-sight velocities of dark matter. For all methods, we plotted the curves corresponding to the optimal value of the initial smoothing scale. The skewers obtained in all cases are consistent with one another.}
\label{fig:skewers}
\end{figure*}

We run 1D-IMS for different values of the initial smoothing length $\lambda_G$.  The results are discussed
in the next section.

\section{Validation of Iteratively Matched Statistics}
\label{sec:validation}

After implementing the methods described in the previous sections, we
assess the accuracy with which we can reproduce the results of the
hydrodynamic simulation. In \S~\ref{sec:skewers} we compare the
performance of the various techniques in reproducing the skewers of
the simulation. In \S~\ref{sec:comparison} and \S~\ref{sec:accuracy_smoothing} we investigate how
accurately the statistics of flux are recovered.

\subsection{Skewers}
\label{sec:skewers}

In Figure \ref{fig:skewers} we show different quantities along one skewer as an example. From top to bottom, we plot the baryon density fluctuations, temperature, velocity field, flux in real space and flux in redshift space. In all panels, solid green lines refer to the skewers extracted from the hydrodynamic simulation, solid blue lines to the ones obtained through GS+FGPA, and solid cyan and dashed red lines to 3D-IMS and 1D-IMS, respectively. Each curve corresponds to the optimal smoothing length for the respective method. The dashed blue line in the third panel from the top
refers to the line-of-sight velocities of DM, Gaussian-smoothed at $\lambda_G=228\,\rm ckpc$. This is the velocity field used in all approximate methods to compute all quantities above (see appendix \ref{app:smoothing_vel}).

We recall that 1D-IMS has 3D-IMS as starting point, and differs from it for two additional transformations to match the dimensionless 1DPS and the PDF of the flux in redshift space with the results from the hydrodynamic simulation. Therefore, the flux in real space, and consequently baryon density fluctuations and temperature fields, are the same as in 3D-IMS. 

We can see that all methods result in skewers that trace those of the
hydrodynamic simulation very well, and are also consistent with one
another. This means that not only is IMS able to reproduce the
statistics of the \lya forest correctly, but it also generates reasonable
mock skewers. This did not obviously have to be the case. 
For example, the
method LyMAS \citep{Peirani_2014} is designed to match the 1DPS and
PDF of the flux from hydrodynamic simulations as well, but only the
more complex version of LyMAS, which involves two additional
transformations, produces reasonable-looking skewers. 
In our techniques, the mappings guarantee that both statistics and flux skewers are reproduced accurately. 

Furthermore, not only is the flux accurately reproduced, but also the
other quantities plotted in Figure \ref{fig:skewers}. The biggest
difference between IMS methods and GS+FGPA is that the
former better reproduces high and narrow density peaks, like the ones
around $3\,\mathrm{cMpc}$ and
$4\,\mathrm{cMpc}$
in Figure
\ref{fig:skewers}.
	
Conversely, this is not always the case for smaller overdensities, such as the one around $0.7\,\mathrm{cMpc}$ in 
Figure \ref{fig:skewers}, where IMS produces a higher density peak than in the hydro.
Note however that at this location the flux in real space is still much more accurate with our methods, since they are designed to match its 3D power spectrum. 
Small differences in $F_{\mathrm{real}}$ can easily yield large differences in density because of the exponential in equation \eqref{eq:tau_real}.  
Such differences persist also in temperature, which in our context is connected to the density through a pure power law. 
Flux skewers in redshift space 
appear to be more similar among the various methods, since the
convolution of real flux with the velocity field and thermal
broadening tends to smooth out the differences. 

\begin{figure*}
\centering
\includegraphics[width=\textwidth]{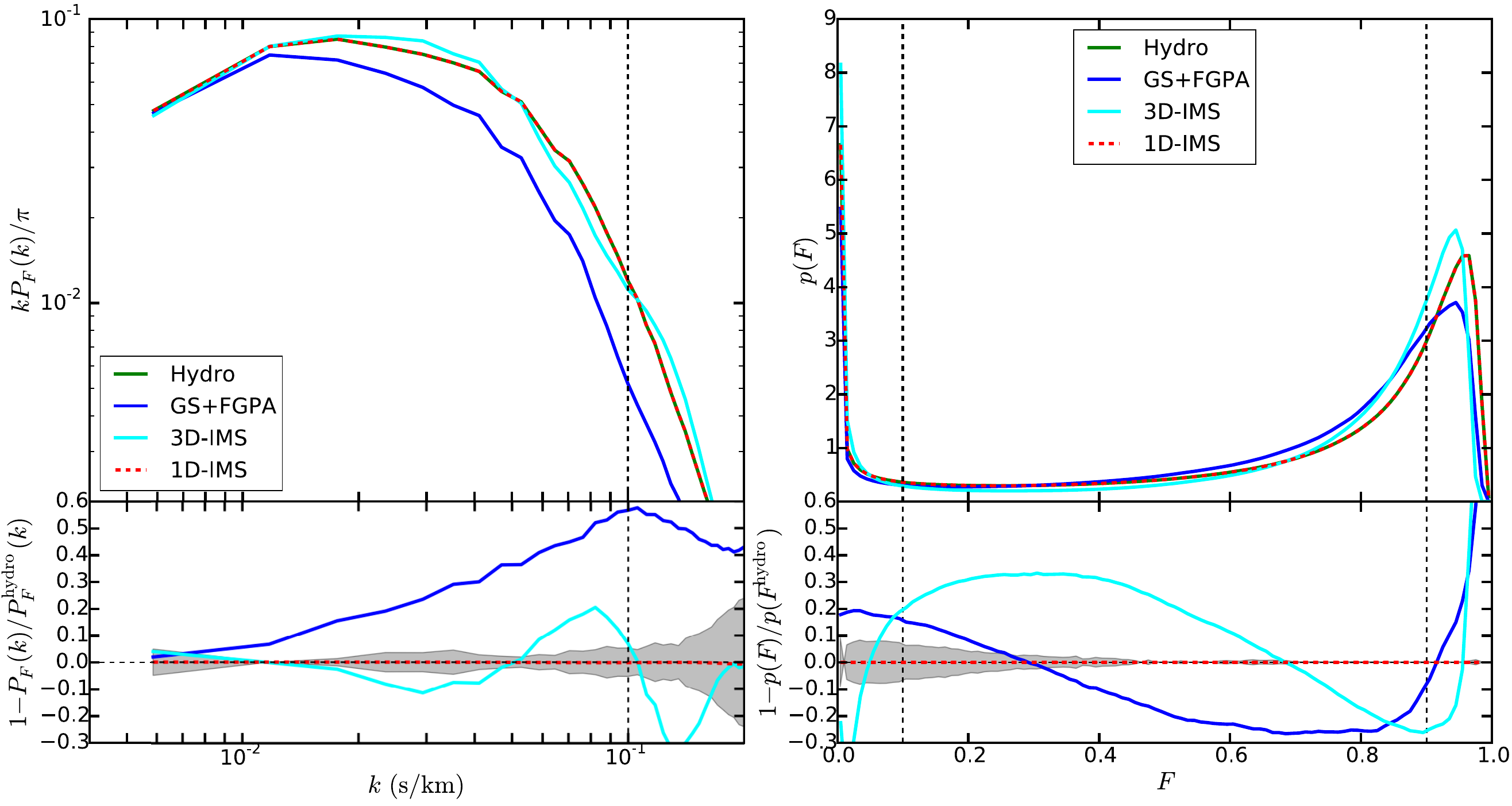}
\caption{\textit{Top panels}: Line-of-sight power spectrum (left) and PDF (right) of flux, given by the reference hydrodynamic simulation (solid green line), GS+FGPA (solid blue line), 3D-IMS (solid cyan line) and 1D-IMS (dashed red line). The results plotted refer to runs with initial smoothing length $228\,\mathrm{ckpc}$. \textit{Bottom panels}: On the left, relative difference between the 1D power spectrum obtained through the different methods tested and the one given by the hydrodynamic simulation. On the right, analogous plot for the PDF. In both panels, the shaded area represents the region within which the relative difference is smaller than the one obtained applying a 1-to-1 temperature density relationship to the baryon density given by the hydrodynamic simulation and using the Gaussian-smoothed line-of-sight velocities of dark matter instead of baryons. In all panels, the dashed vertical lines delimit the dynamic range considered to compute the accuracy. Horizontal dashed black lines mark the zero difference level and are meant to guide the eye. Our methods reproduce the line-of-sight better than GS+FGPA. In particular, 1D-IMS matches both power spectrum and PDF by construction.} 
\label{fig:PS_and_PDF_all}
\end{figure*}

\begin{figure*}
\centering
\includegraphics[width=0.9\textwidth]{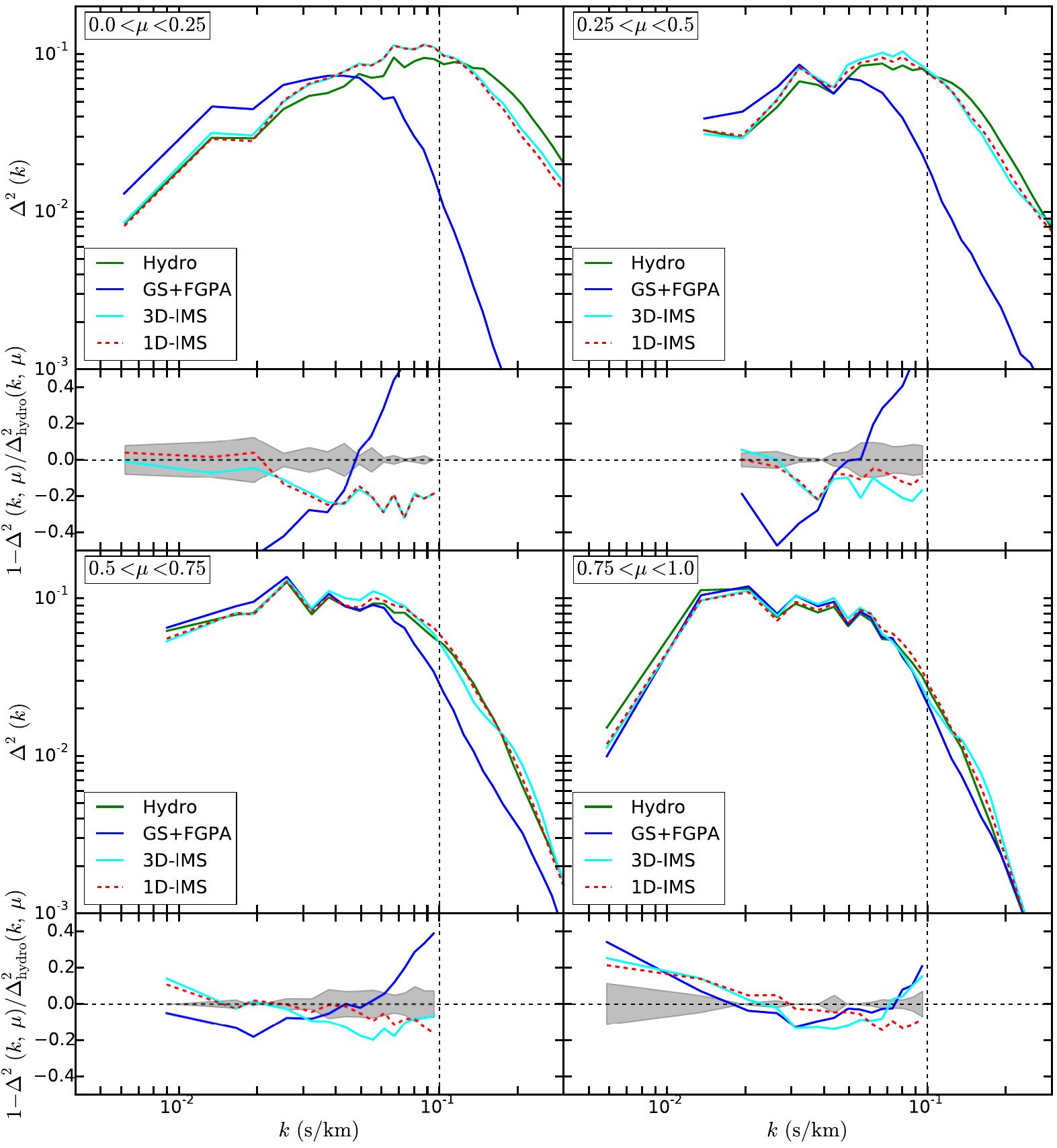}
\caption{We show the dimensionless 3D power spectrum $\Delta^2(k,\,\mu)$ of the flux given by our reference hydrodynamic simulation (solid green lines), by GS+FGPA (solid blue lines), 3D-IMS (solid cyan lines) and 1D-IMS (dashed red lines). The results plotted refer to runs with initial smoothing length $\lambda_G=228\,\mathrm{ckpc}$. We considered 4 bins of $\mu$. For each one of them, there are two panels. The upper one shows  $\Delta^2(k,\,\mu)$ versus $k$ in the $\mu$-bin considered, the lower one the relative difference between the spectra. In all panels, the dashed vertical lines delimit the dynamic range considered to compute the accuracy. Horizontal dashed black lines mark the zero difference level and are meant to guide the eye. Shaded areas represent the regions within which the relative difference is smaller than the one obtained applying a 1-to-1 temperature density relationship to the baryon density given by the hydrodynamic simulation and using the Gaussian-smoothed line-of-sight velocities of dark matter instead of baryons. Our methods perform better than GS+FGPA in all $\mu$-bins.} 
\label{fig:3DPS_all}
\end{figure*}

\begin{figure*}
\centering
\includegraphics[width=\textwidth]{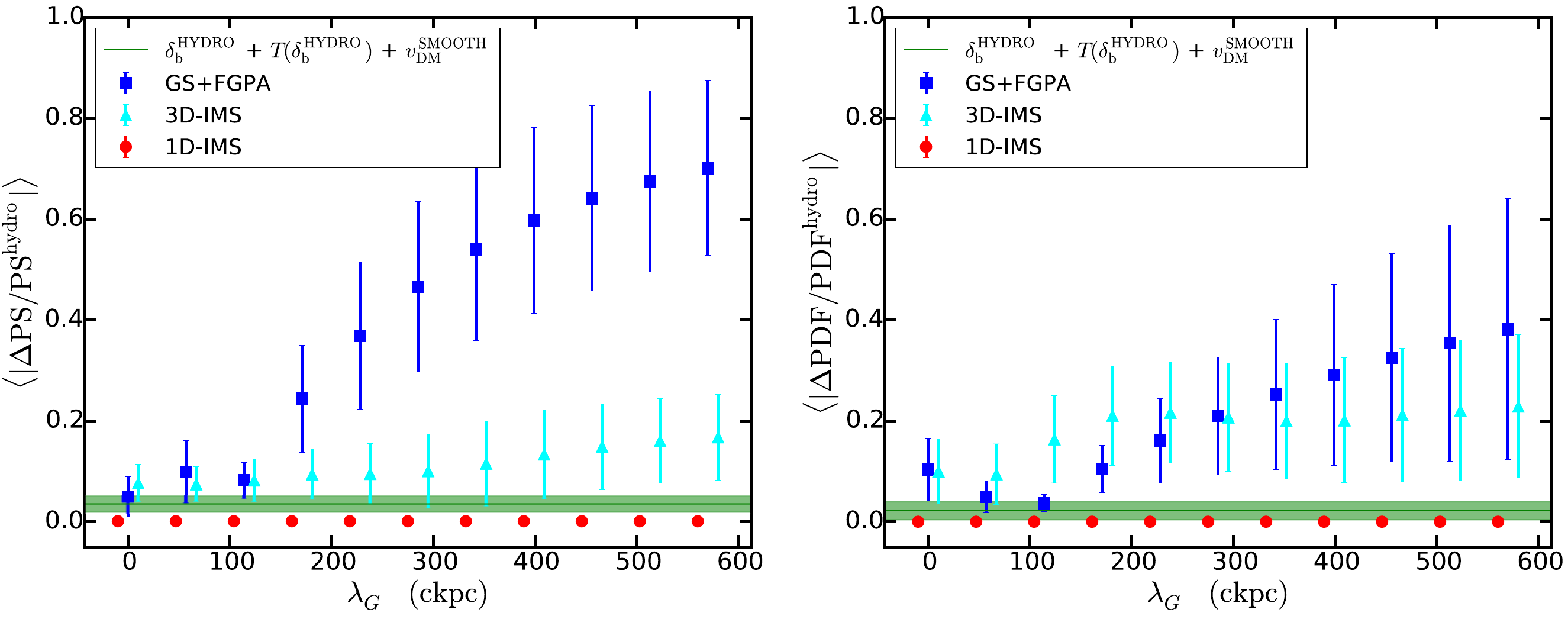}
\caption{Accuracy of the different methods tested in reproducing the flux dimensionless line-of-sight power spectrum (left panel) and PDF (right panel) given by the reference hydrodynamic simulation, as a function of the initial smoothing length $\lambda_G$. Markers indicate the mean values of the accuracy, while error bars represent the root-mean-square of the accuracy in the dynamic ranges considered. Blue squares refer to GS+FGPA, cyan triangles to the 3D-IMS and red circles to the 1D-IMS. An offset of $\pm10\,\mathrm{ckpc}$ has been applied to 3D-IMS and 1D-IMS markers to make the plot more readable. The horizontal green line shows the mean accuracy obtained by applying a 1-to-1 temperature-density relationship to the baryon density field and using the Gaussian-smoothed line-of-sight velocities of dark matter baryons. The green band represents the root-mean-square of the accuracy in this case. Our methods are overall more accurate and less dependent on the initial smoothing scale than GS+FGPA.} 
\label{fig:accuracy_1D}

\centering
\includegraphics[width=\textwidth]{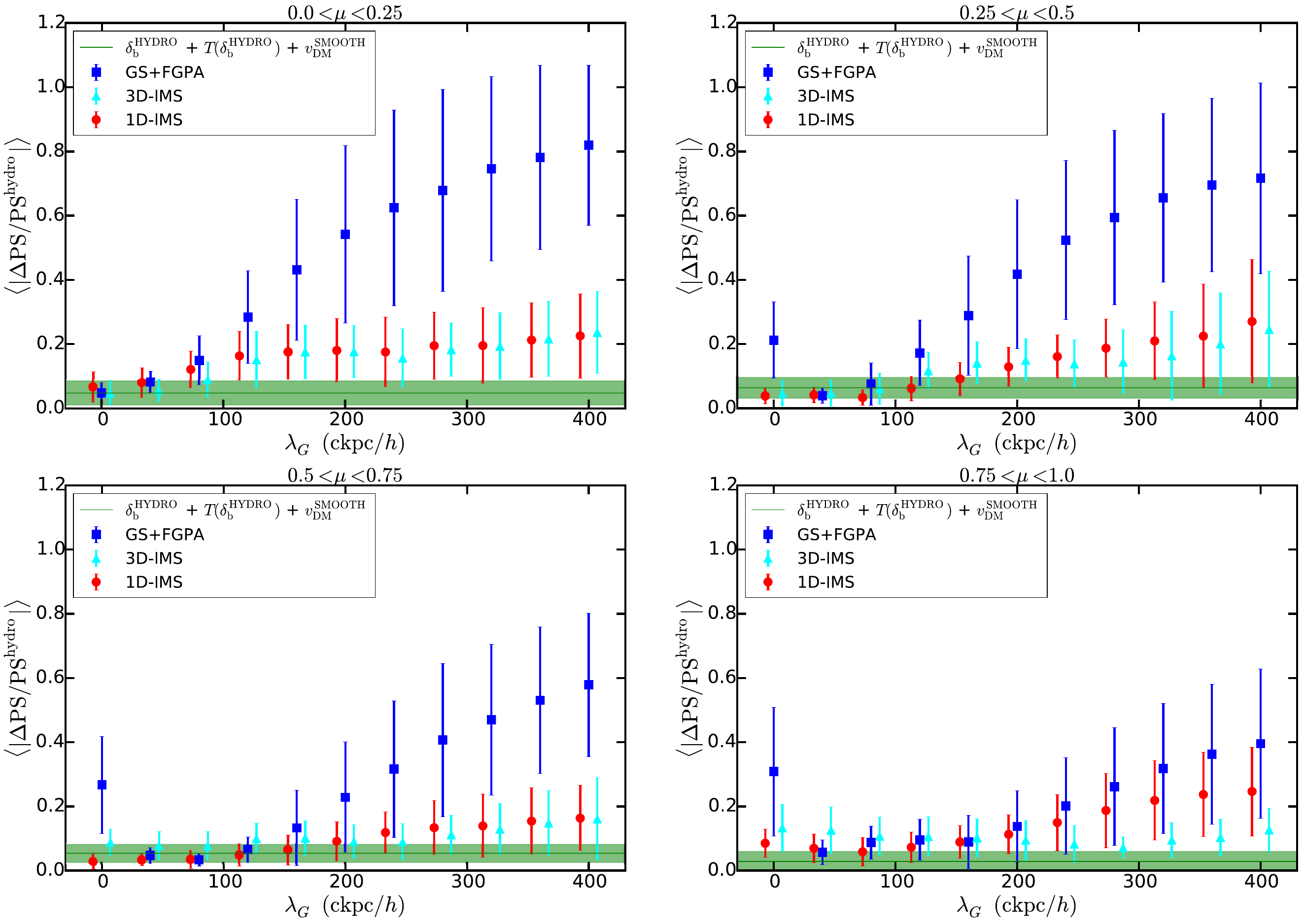}
\caption{Accuracy of the different methods tested in reproducing the dimensionless 3D power spectrum $\Delta^2(k,\,\mu)$ of the flux given by the reference hydrodynamic simulation, as a function of the initial smoothing length $\lambda_G$. Each panel shows the results obtained for a different bin of $\mu$. In all panels, markers indicate the mean values of the accuracy, while error bars represent the root-mean-square of the accuracy in the dynamic ranges considered. Blue squares refer to GS+FGPA, cyan triangles to the 3D-IMS and red circles to the 1D-IMS. An offset of $\pm10\,\mathrm{ckpc}$ has been applied to 3D-IMS and 1D-IMS markers to make the plot more readable. The horizontal green lines show the mean accuracy obtained by applying a 1-to-1 temperature-density relationship to the baryon density field and using the Gaussian-smoothed line-of-sight velocities of dark matter instead of baryons. The green bands represent the root-mean-square of the accuracy in this case. In all $\mu$-bins, our methods are overall more accurate and less dependent on the initial smoothing scale than GS+FGPA.} 
\label{fig:accuracy_3DPS}
\end{figure*}

\subsection{Comparison of the Methods}
\label{sec:comparison}

All methods we considered have GS+FGPA as their starting point, with $\lambda_G$ as a free parameter. We now compare the flux statistics given by each method with the ones from the hydrodynamic simulation, varying $\lambda_G$ in the range $0-570 \, \mathrm{ckpc}$, in steps of $57\,\mathrm{ckpc}$. The dependence on this parameter of the accuracy in reproducing the various statistics is different for each method. It is generally possible to identify an optimal value of $\lambda_G$ for a given method at matching a certain statistic, but this may not be optimal for all flux statistics.  

In the top panels of Figure \ref{fig:PS_and_PDF_all} we compare 
1DPS and PDF of the flux in redshift space given by the hydrodynamic simulation (solid green line) to the approximate methods considered. Solid blue, solid
cyan and dashed red lines refer to GS+FGPA,
3D-IMS and 1D-IMS, respectively. We plotted the curves corresponding to $\lambda_G=228\,\rm ckpc$ for each method. Since 1D-IMS is
designed to match dimensionless 1DPS and PDF of the flux, solid green and dashed red
lines are indistinguishable. 
We can see that 3D-IMS reproduces well both 1DPS and PDF, whereas GS+FGPA does not recover well the 1DPS. The lower panels make the comparison
quantitative, showing the relative error of each method at recovering
the results of the reference simulation. The gray shaded area represents
the region within which the relative difference is smaller than the
one obtained applying the FGPA to the baryon density given by the
hydrodynamic simulation and using the smoothed DM velocity field, as
explained in section \ref{sec:FGPA_accuracy}. We recall that this sets
the limits on the accuracy due to adopting the DM-smoothed velocity
field and neglecting the scatter in the temperature-density
relationship of the IGM.

Figure \ref{fig:PS_and_PDF_all} then tells us that 1D-IMS is able to recover the information lost with these approximations by construction, since it was forced to match the redshift space 1DPS and PDF of the hydrodynamic simulation. 
In contrast, the flux PDF given by 3D-IMS does not appear very
accurate, perhaps even erroneously suggesting a flaw in the method. This
is not the case, as 3D-IMS matches the PDF of the flux in real space,
whereas in the right panel we are considering the PDF of the flux in
redshift space. Although the relative error of the 3D-IMS PDF 
is as large as 30\% at $F=0.2$, 
the average accuracy is 15\%. When the optimal value of $\lambda_G$ is used for 3D-IMS ($57 \, \rm ckpc$), the PDF is reproduced with an average accuracy of 8\%. The variability of the accuracy at different flux values is not too surprising, because in the last step of 3D-IMS we smooth the pseudo baryon density field to remove hot pixels and this impacts the accuracy of the corresponding flux PDF.

Figure \ref{fig:3DPS_all} shows the 3DPS given by the simulation and
the various methods at $\lambda_G=228\,\rm ckpc$, 
as well as the
relative error in matching this statistic. The color coding is the
same as in Figure \ref{fig:PS_and_PDF_all}. Each panel refers to a
different $\mu$-bin of the 3DPS. In all bins, the accuracy of 3D-IMS and 1D-IMS looks on average comparable to the limit set by the FGPA.  

In Figure \ref{fig:3DPS_all} one can clearly see that GS+FGPA in the top-left panel ($0.0 < \mu < 0.25$, farthest from the line-of-sight) 
does not match the hydrodynamic result as well as in the other panels. This
is due to the different effects at work at different directions from
the line-of-sight. For very transverse modes $(0.0 < \mu < 0.25)$
the behavior of baryons is mostly influenced by the filtering scale. 
Whereas for modes that are parallel to the line-of-sight $(0.75 < \mu < 1.0)$
the effect of the Jeans filtering is degenerate with thermal broadening and redshift space distortions
\citep{Rorai_2013}. As a result, in the bin closest to the line-of-sight  $(0.75 < \mu < 1.0)$ one can compensate a bad choice of $\lambda_G$ with an accurate
description of the thermal state of the IGM, whose parameters ($T_0$
and $\gamma$) are obtained by fitting outputs of the hydrodynamic
simulation. But it is not possible to apply this correction in
the bin with modes most transverse to the line-of-sight, where the effect of the
filtering scale dominates the shape of the 3D power. 
In all $\mu$-bins, both 3D-IMS and 1D-IMS present an accuracy between 4\% and 30\%, depending on the smoothing scale considered. 
Furthermore, 3D-IMS is superior to GS+FGPA to the extent that it also matches the dimensionless 3D power spectrum of flux in real space by construction.

Regarding 1D-IMS, it might seem puzzling that it does not perfectly
match the result of the hydrodynamic simulation in the bin closest to
the line-of-sight, since 1D-IMS is forced to reproduce the
dimensionless line-of-sight power spectrum by construction. However, the 3DPS
in the bin closest to the line-of-sight is not exactly the
1DPS. Indeed, the power spectrum in that bin considers all flux
fluctuations whose wavevector forms an angle with the line-of-sight
such that its cosine is between 0.75 and 1. This is a 3D region in
redshift space. In the case of the 1DPS, the situation is much
different, since one considers flux fluctuations exclusively along the
direction of the line-of-sight. Therefore, the 1DPS and 3D power for the bin closest to
the line-of-sight 
are not exactly the same, and matching the 1DPS by construction does not guarantee a perfect
match in this bin of the 3DPS. It is true, however, that the agreement should be much better if one considers $\mu$ values
progressively closer to being parallel to the line-of-sight ($\mu=1$), which we have verified directly. 

\subsection{Accuracy versus Smoothing Length}
\label{sec:accuracy_smoothing}

The best simulations reproducing BOSS/DESI-like surveys have a mean inter-particle separation of $\sim 400 \,\rm ckpc$. As we have already mentioned, Gaussian-smoothing below the the inter-particle separation has a negligible effect. Therefore, it is of great interest to test the accuracy of GS+FGPA and our methods at different values of the smoothing length, including $\lambda_G > 300 \, \rm ckpc$.  
For this purpose, we compute the mean and
root-mean-square of the accuracy for each value of $\lambda_G$, as
explained in section \ref{sec:FGPA_accuracy}.

We show the results of
our analysis for the 1DPS and PDF in Figure \ref{fig:accuracy_1D}, in
the left and right panels, respectively. The information given by
Figure \ref{fig:PS_and_PDF_all} is here condensed in three points, one
for each method, at the corresponding value of $\lambda_G$. Blue
squares represent the mean accuracy of GS+FGPA
and the corresponding error bars the root-mean-square. Likewise, cyan
triangles and red circles refer to 3D-IMS and 1D-IMS,
respectively. The information encoded by the gray shaded areas in
Figure~\ref{fig:PS_and_PDF_all}, which shows the limitations of the
FGPA, is represented by the green band in Figure
\ref{fig:accuracy_1D}. Hence, the green line shows the mean accuracy
given by the FGPA implemented as in section \ref{sec:FGPA_accuracy}
and the shaded green area delimits 1-$\sigma$ deviations from this
mean. There is no dependence on $\lambda_G$ in this case, because the
flux within the FGPA is computed from the baryon density field given
by the hydrodynamic simulation and not from a Gaussian-smoothed DM
density field. Figure \ref{fig:accuracy_3DPS} shows the results for
the 3DPS, with the same format as Figure
\ref{fig:accuracy_1D}. Each panel of Figure \ref{fig:accuracy_3DPS}
refers to a different $\mu$-bin.

As expected, Figures \ref{fig:accuracy_1D} and \ref{fig:accuracy_3DPS} show that GS+FGPA is strongly dependent on $\lambda_G$. The optimal value appears to be around $114 \,
\mathrm{ckpc}$ for the 1DPS and between $57 \,\mathrm{ckpc}$ and $114\,
\mathrm{ckpc}$ for the PDF. Around these values, the mean accuracy is
~7\% and ~4\% for the 1DPS and PDF, respectively, as can be seen
from the blue points in Figure \ref{fig:accuracy_1D}.\footnote{The minimum in
the accuracy of the 1DPS at $\lambda_G=0\,\mathrm{ckpc}$ would suggest
that the best result is obtained without smoothing the DM density at
all. If we apply no smoothing, we are actually limited by the
resolution of the simulation. In our context, the DM was solved using a PM code on a grid with size of $28\,\mathrm{ckpc}$ and that also corresponds to the inter-particle separation. The DM density was also implicitly smoothed by the CIC kernel on that scale, which is thus the effective smoothing length corresponding to $\lambda_G=0$. 
There is hence nothing peculiar about the
point at $\lambda_G=0$.
Furthermore, the overall accuracy
corresponding to this value is actually similar to the value obtained at
$\lambda_G=57\,\mathrm{ckpc}$.}

The trend of the accuracy of GS+FGPA in reproducing the 3DPS is similar to the one of the 1DPS (blue squares in Figure \ref{fig:accuracy_3DPS}). 
The mean accuracy achieved in the different $\mu$ bins at the optimal scale
for the 3DPS ($\lambda_G = 57 \mathrm{ckpc}$)
is around 4\%. Remarkably, the accuracy of GS+FGPA for all
statistics approaches the limit set by the FGPA, as long as the
``correct'' smoothing length is chosen. Since the optimal scales for
the statistics considered vary up to a factor of two, one should decide
in advance whether to prioritize 1DPS, 3DPS or PDF. For $\lambda_G \gtrsim 171\, \rm ckpc$, the accuracy of all statistics gets worse than $\sim 20\%$.
Moreover, the error bars are very
large for smoothing scales $\gtrsim 200 \, \mathrm{ckpc}$. 
As
such, it can be much worse than the mean in certain ranges of
$k$-modes and flux. 
We also note that the performance of GS+FGPA 
degenerates as one moves farther from the line-of-sight, as previously discussed in the
context of Figure~\ref{fig:3DPS_all}.

Even for initial smoothing lengths $\gtrsim 200 \, \mathrm{ckpc}$,
3D-IMS results in an accuracy better than 20\% for 1DPS and PDF, as shown
by the cyan triangles in Figure \ref{fig:accuracy_1D}, performing
significantly better than the Gaussian method for these large
smoothing lengths.  At smaller smoothing lengths, 3D-IMS is basically
as accurate as GS+FGPA. The accuracy in the
1DPS is better than in the PDF. This is not so surprising since, as we
already pointed out, we ended our iterations matching the PDF of the
flux in real space, and not redshift space, of the hydrodynamic
simulation. 
Figure \ref{fig:accuracy_3DPS} shows that 3D-IMS
does a remarkable job of reproducing the 3DPS, with an accuracy
comparable to the FGPA at small $\lambda_G$, 
and still around 7\% even for initial smoothing lengths as large as 500
$\mathrm{ckpc}$. Moreover, the accuracy of 3D-IMS is only weakly dependent on $\lambda_G$, and
it performs much better than GS+FGPA
for large smoothing lengths.

By construction 1D-IMS matches the 1DPS and PDF resulting in an
accuracy of 0.03\% independent of $\lambda_G$ (red circles in Figure
\ref{fig:accuracy_1D}). Figure \ref{fig:accuracy_3DPS} shows that
1D-IMS preserves 3D correlations, yielding an accuracy in the
3DPS of 3.3\% in the best case ($57\,\mathrm{ckpc}$ in the bin $0.25 < \mu < 0.5$) and 27\% in the worst one ($570\,\mathrm{ckpc}$ in the bin $0.25 < \mu < 0.5$). In all bins, 1D-IMS is as accurate as Gaussian
smoothing at small smoothing lengths and it performs better than this method for $\lambda_G \gtrsim 142\,\mathrm{ckpc}$. 

The accuracy of 1D-IMS improves as the $\mu$-bin considered approaches the
line-of-sight. 
It performs worse than 3D-IMS in the bin closest to it $(0.75 < \mu < 1.0)$. This is
counter-intuitive, but we recall that the most parallel bin takes into account
correlations in a 3D region of space and is thus conceptually distinct
from the 1DPS, which 1D-IMS matches by construction. When recovering
the 3DPS in the bin closest to the line-of-sight $(0.75 < \mu < 1.0)$,
it is still more important to correctly reproduce the 3D correlations
rather than the correlations along the line-of-sight. This is why
3D-IMS looks better than 1D-IMS close to the line-of-sight. Similar to 3D-IMS, 
the accuracy of 1D-IMS is only
weakly dependent on the smoothing length, and  much better than GS+FGPA
for large smoothing lengths.

Among the methods considered in this work, 1D-IMS seems to perform
the best. Indeed, it perfectly matches the 1DPS and PDF (by construction)
and reproduces the 3DPS with a good accuracy. If one is
is primarily interested in the 3DPS, 3D-IMS may be more suitable,
since it yields the best
accuracy in this statistic, although the differences with 1D-IMS are small.
The drawback of 3D-IMS is the relative inaccuracy in the
1DPS and PDF compared to 1D-IMS, which matches these statistics by construction. 
The Gaussian
smoothing can recover all statistics as well as 3D-IMS, 
provided the appropriate
$\lambda_G$ is adopted.  In particular, the errors in estimating the 3DPS in the bin farthest from the line-of-sight ($0.0 < \mu < 0.25$, top-left panel in Figure \ref{fig:accuracy_3DPS}) are larger than $\sim20\%$ for
$\lambda_G\gtrsim 171\,\mathrm{ckpc}$.
For comparison, 3D-IMS and 1D-IMS achieve $\lesssim$10\% accuracy for $\lambda_G\lesssim228\,\mathrm{ckpc}$ in the aforementioned $\mu$-bin. 
This means that our methods are able to recover information that gets
otherwise lost when performing a Gaussian smoothing. They are accurate and
computationally cheap ways to reproduce the statistics of the \lya
forest, which have promise for future modeling and data analysis. 

We have applied the same analysis described so far also to two snapshots at redshifts $z=2$ and $z=4$, respectively. The accuracy of all methods are comparable with the results obtained at $z=3$, meaning that the techniques tested are robust in the range $2<z<4$. We have also verified that, with $256^3$ resolution elements, the accuracy of all methods is very close to the values obtained for our reference simulation. This means that the accuracy of the methods has converged in our study. 

When applying our methods, the choice of the initial smoothing length for the DM density is set by the inter-particle separation of the simulation adopted. If this is smaller than the optimal smoothing length, then one should smooth the DM density at the optimal $\lambda_G$. Otherwise, the best one can do is adopting a smoothing length of the order of the inter-particle separation. In any case, the smoothing scale for the DM line-of-sight velocities can be larger than the value adopted for the DM density. Indeed, the velocity field itself is smooth in voids \citep{van_de_Weygaert_1993, Argon-Calvo_2013} and these are the most relevant regions for our study, as the exponentiation in equation \eqref{eq:tau_real} suppresses large overdensities. In our analysis, we kept the smoothing length for the DM velocity
field fixed to $228\,\rm ckpc$, which is the value that yields the
best overall accuracy in reproducing the flux statistics considered
(see appendix \ref{app:smoothing_vel}). In this way, we focused on the impact of the initial smoothing of the DM density field
on the accuracy of the methods. Due to our choice of
optimizing the smoothing length of the DM velocity field, the errors
quoted for the different methods are minimized. However, even if we
did not use the optimal $\lambda_G$ for the velocity, the trend of the
accuracy versus the smoothing length of the DM density field would be
unaffected, as well as the the rank ordering of the accuracy of the
various techniques investigated (see appendix \ref{app:smoothing_vel}
for a detailed discussion).

\begin{figure*}
\centering
\includegraphics[width=\textwidth]{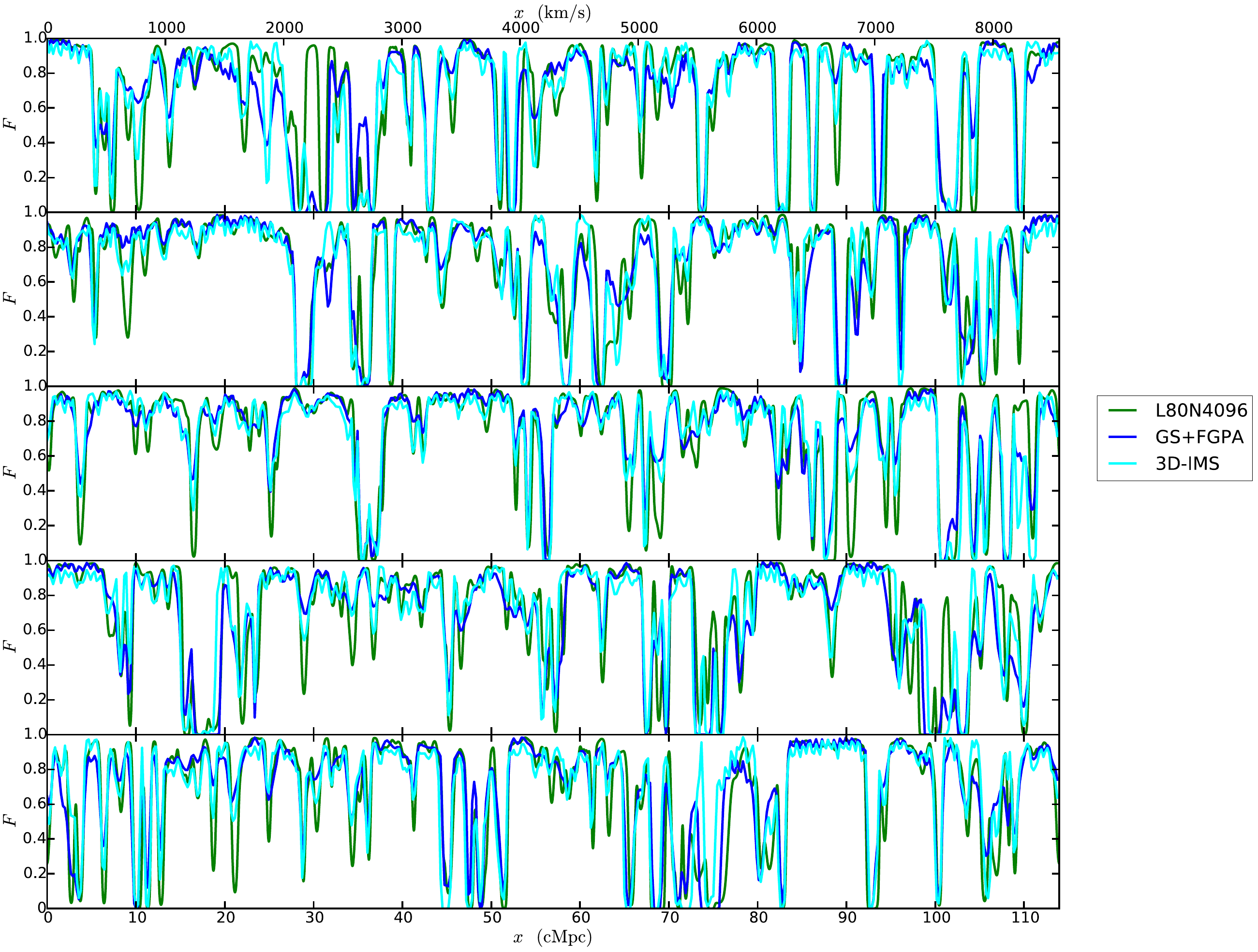}
\caption{Sample of five flux skewers obtained through different methods: L80N4096 hydrodynamic simulation with box size $114\, \rm cMpc$ and $4096^3$ resolution elements (solid green line), which we assume to be the ``truth'', GS+FGPA with smoothing length $228\,\rm cMpc$ (solid blue line) and 3D-IMS (solid cyan line). The skewers obtained through all methods are consistent with one another.}
\label{fig:skewers_bigbox}
\end{figure*}

\begin{figure*}
\centering
\includegraphics[width=\textwidth]{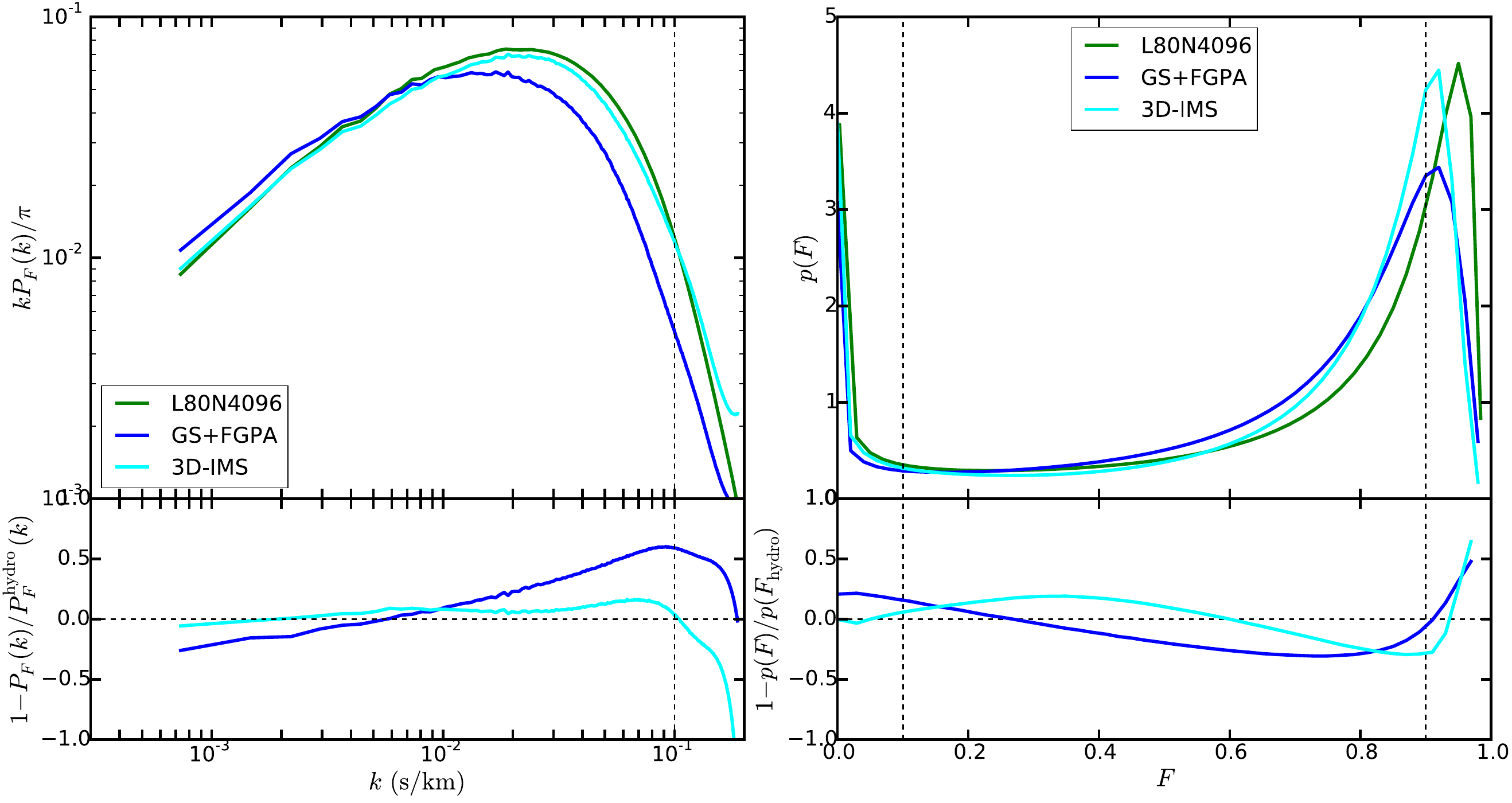}
\caption{Flux line-of-sight power spectrum (left) and PDF (right) given by L80N4096 hydrodynamic simulation (solid green line), assumed to be the ``truth'', and  obtained applying GS+FGPA (solid blue line) and 3D-IMS (solid cyan line) to the DM-only simulation. In all panels, the dashed vertical lines delimit the dynamic range considered to compute the accuracy. All results plotted refer to runs with initial smoothing length $228\,\mathrm{ckpc}$. This is the smallest smoothing allowed by the resolution of the simulation, so the Gaussian smoothing is already optimized. Nevertheless, it is very inaccurate in recovering the line-of-sight power spectrum, meaning that 3D-IMS is certainly superior. }
\label{fig:DM_1DPS_PDF}
\end{figure*}

\begin{figure*}
\centering
\includegraphics[width=\textwidth]{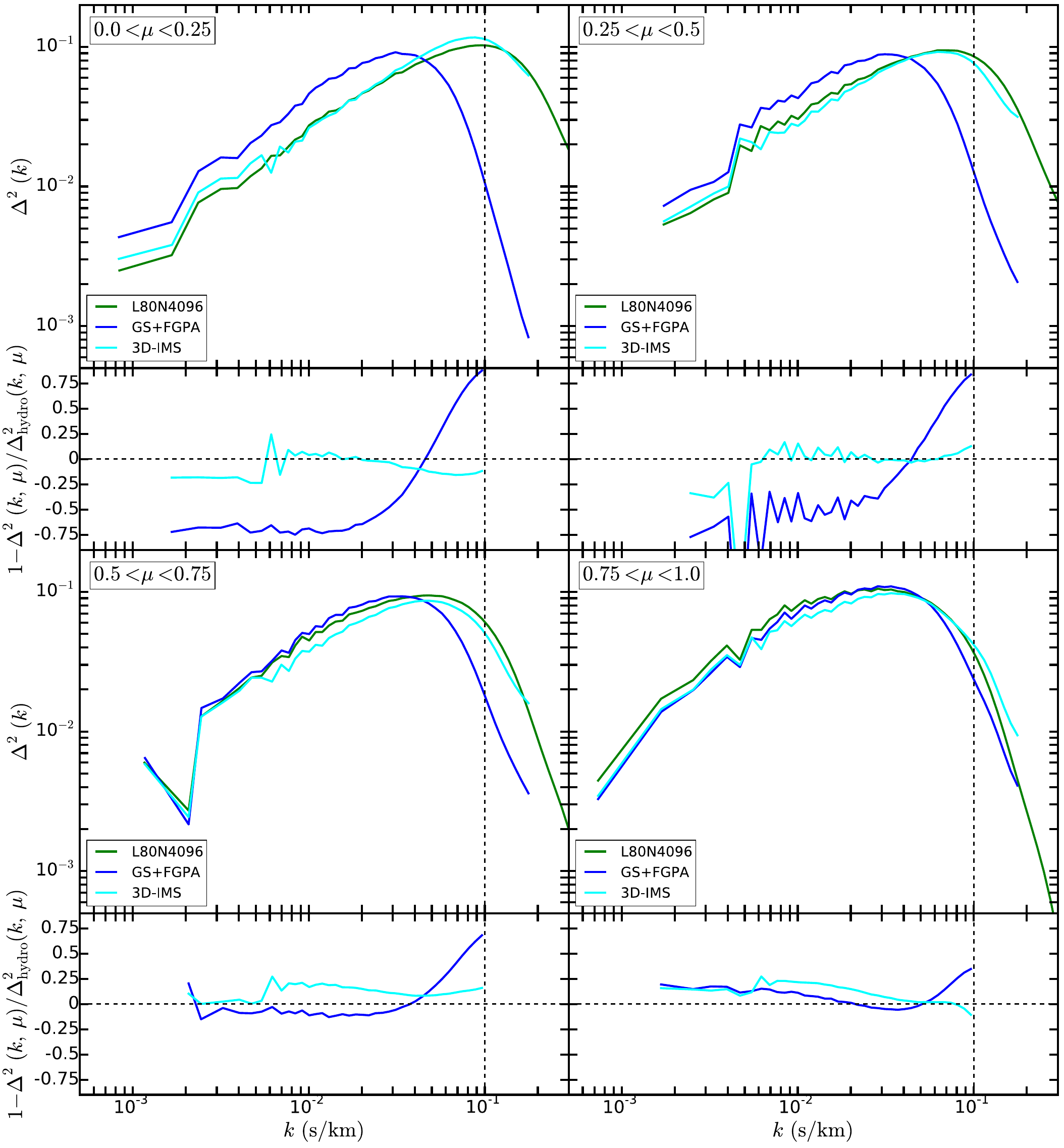}
\caption{Dimensionless 3D power spectrum $\Delta^2(k,\,\mu)$ of the flux given by L80N4096 hydrodynamic simulation (solid green lines), assumed to be the ``truth'', and obtained applying GS+FGPA (solid blue lines) and 3D-IMS (solid cyan lines) to the DM-only simulation. The results plotted refer to runs with initial smoothing length $\lambda_G=228\,\mathrm{ckpc}$. We considered 4 bins of $\mu$. Each panel shows $\Delta^2(k,\,\mu)$ versus $k$ in the $\mu$-bin considered. In all panels, the dashed vertical lines delimit the dynamic range considered to compute the accuracy. Whereas 3D-IMS is accurate and its performance does not depend strongly on the $\mu$-bin considered, GS+FGPA fails at reproducing the true $\Delta^2(k,\,\mu)$ in the $\mu$-bins closer to the line-of-sight ($0.0<\mu<0.25$ and $0.25<\mu<0.5$).}
\label{fig:DM_3DPS}
\end{figure*}

\section{Large-Volume Collisionless Simulation}
\label{sec:DM_application}

We want to check if our methods still perform well when applied to an actual N-body run, with a larger box than our calibrating hydrodynamic simulation. In fact, in the previous sections we have validated our methods extracting all relevant fields from the same hydrodynamic simulation. However, the purpose of approximate methods is avoiding expensive hydrodynamic simulations. In practice, one would assume a certain model for the flux statistics and apply our techniques to a large-box low-resolution DM-only run. In this way, one would be able to probe large scales and at the same time accurately describe the small-scale physics thanks to our iterative procedure.

We consider the snapshot at redshift $z=3$ of a Gadget DM-only run with a box size of $114\,\rm cMpc$ and $512^3$ particles. We CIC-bin the particle positions and velocities on a grid with $512^3$ elements, to get the density and velocity fields. To mock baryonic pressure, we smooth both fields with a length scale of $\lambda_G=228\,\rm ckpc$, very close to the cell size ($223\,\rm ckpc$). This is the smallest smoothing length one can choose to have a non-negligible effect on the density and velocity fields. \footnote{The smoothing scale adopted here is also very close to the optimal value for the velocity field determined when validating our method with the smaller hydrodynamic simulation (see appendix \ref{app:smoothing_vel}).} We then apply GS+FGPA in \S~\ref{sec:Gauss_smoothing}.

To apply 3D-IMS, we need an input model for the 3D power spectrum and PDF of the flux in real space. These statistics are computed from the flux in real space given by our hydrodynamic simulation. However, its box size is smaller than the one of the N-body simulation. This does not affect the computation of the PDF, but it poses some problems with the 3D power spectrum, as the hydrodynamic simulation lacks the large modes which are present in the DM-only simulation. To generalize the 3D-IMS method, we construct the transfer function defined in equation \eqref{eq:transfer} as follows. First of all, we fit the 3D power spectrum of the flux in real space given by the hydrodynamic simulation with the formula provided by \cite{Kulkarni_2015}. Then, at every iteration of 3D-IMS, we define the transfer function applying equation \eqref{eq:transfer} for $k$-modes larger than the fundamental mode $k_{\rm f}^{\rm HYDRO}$ of the hydrodynamic simulation. For the modes smaller than $k_{\rm f}^{\rm HYDRO}$ we set the transfer function to a constant, equal to the value assumed at $k_{\rm f}^{\rm HYDRO}$. In this way, we have a continuous transfer function, which simply rescales by a constant the large-scale modes probed only by the DM-only simulation. This guarantees that any peculiar large-scale feature in the DM-only simulation (e.g. BAO signal), will not be affected. 

Modeling the 1DPS of the flux in redshift space presents similar issues. Once again, one should come up with a method to estimate the large-scale Fourier modes without actually running a calibrating large-box hydrodynamic simulation. It would then be desirable to adopt an approach analogous to the one described for the 3D power spectrum of the real flux field. Unfortunately, there is no simple fitting function available for the 1DPS of the flux in redshift space which would grant the high level of accuracy we are aiming for. Therefore, we are not applying 1D-IMS to the DM-only simulation. Nevertheless, following \cite{Kulkarni_2015}, future work may provide a fitting function for the flux 1DPS as well.

To assess the accuracy of the various techniques in this test, we shall compare the results of each method with the flux statistics obtained from a Nyx hydrodynamic run with a box size of $114\, \rm cMpc$ and $4096^3$ resolution elements, which we assume to be the ``truth''. This simulation, to which we shall refer as ``L80N4096'', covers the largest modes present in the DM-only simulation and, at the same time, has the same resolution limit ($\sim 28 \, \rm ckpc$) as the small calibrating hydrodynamic simulation, thus resolving the Jeans scale.

In Figure \ref{fig:skewers_bigbox} we show a sample of five skewers extracted from the hydrodynamic simulation (solid green line) and obtained through GS+FGPA and 3D-IMS (solid blue and cyan lines, respectively). By visual inspection, all skewers look consistent with one another. 

In the upper-left panel of Figure \ref{fig:DM_1DPS_PDF} we show the 1DPS of the flux given by the hydrodynamic simulation (solid green line) and obtained applying GS+FGPA and 3D-IMS to the DM-only run (solid blue and cyan lines, respectively). In the lower-left panel, we plot the relative difference of the flux 1DPS obtained in each case, with respect to the results of L80N4096. In the right panels, we show analogous plots for the flux PDF, following the same color coding. Applying the same analysis outlined in previous sections, we find out that the average and root-mean-square of the accuracy with which the flux 1DPS is recovered are 41\% (10\%) and 18\% (4\%) for GS+FGPA (3D-IMS), respectively. Thus, GS+FGPA is not accurate at all in this context. This fact should be born in mind when dealing with low-resolution DM-only simulation with $\sim 100 \, \rm cMpc$ boxes. On the contrary, the accuracy is dramatically improved by 3D-IMS. For the PDF, the mean accuracy and its root-mean-square are 18\% (13\%) and 10\% (6\%) for GS+FGPA (3D-IMS), respectively. Therefore, also this statistics is better reproduced by 3D-IMS. We note that the precision achieved for the flux 1DPS and PDF is of the same order of what we obtained when we extracted the DM density field from the same hydrodynamic simulation used for the calibration. 

In Figure \ref{fig:DM_3DPS} we plot the 3DPS of the flux given by the hydrodynamic simulation and obtained with GS+FGPA and 3D-IMS, applied to the DM-only simulation. The color coding is the same as in Figure \ref{fig:DM_1DPS_PDF}. The different panels show the 3DPS in four evenly  spaced $\mu$-bins. We recall that the bin $0.0 < \mu < 0.25$ corresponds to modes farther from the line-of-sight, whereas the bin $0.75 < \mu < 1.0$ is the closest to it. Below the plots obtained for each bin, we show the relative difference of the results of each method with respect to the ones given by L80N4096. Once again, we applied the same analysis technique adopted throughout this work, obtaining that the mean accuracy of 3D-IMS is $\sim$10\% in all $\mu$-bins. On the contrary, the accuracy of GS+FGPA is strongly dependent on the  $\mu$-bin considered. The accuracy is 10\% in the $\mu$-bin farthest from the line-of-sight, degrading up to 58\% in the bin closest to it. These results are consistent to the findings presented in the previous sections.

The accuracy of 3D-IMS in reproducing the 1DPS, 3DPS and PDF of the flux in redshift space is higher than in the case of GS+FGPA. The accuracy is comparable to the results obtained when applying 3D-IMS to the DM field extracted from our reference simulation. To probe the limitations of our technique, we applied the analysis explained in the present section also to the $256^3$ Gadget run, adopting the same calibrating simulation. We verified that the ratio of the accuracy of the two methods does not change significantly. In conclusion, our method is solid when applied to a large-box low-resolution DM-only simulation. This achievement makes our method attractive for studies requiring both large volumes and high resolution simulations, for which running hydrodynamic simulations is not a viable option. One example is modeling the signature of the BAO on the \lya forest flux power spectrum. 

The results presented in this section clearly show that GS+FGPA can yield a very poor accuracy with respect to what would be obtained through a hydrodynamic simulation. Any result claimed after applying this technique should then be considered carefully. Future works making use of GS+FGPA should refer to Figures \ref{fig:accuracy_1D} and \ref{fig:accuracy_3DPS} to assess the error intrinsic in the method adopted. 

1D-IMS has not been validated for a large-box DM-only run because of the lack of a recipe to model the flux 1DPS (e.g. an analytic fitting formula), extending it to large scales. Though, we are confident that in future work such fitting procedure could be provided. If such technique becomes available, we do not expect that 1D-IMS will fail the test presented in this section. Indeed, 1D-IMS and 3D-IMS are both grounded on the Iteratively Matched Statistics technique. We proved that 3D-IMS is accurate when applied to a large-box DM-only run, so this is encouraging for 1D-IMS as well.

\section{Comparison to Previous Works}
\label{sec:comparison_past}

With our analysis technique, we defined a criterion to assess the accuracy in reproducing 1DPS, PDF and 3DPS of flux skewers, which we wish will be used also by other authors in the future. This would make the comparison with upcoming works more direct and straightforward. It is of course interesting to compare our results with previous work. We do that at the best of our possibilities, since the statistics discussed in the relevant literature are not always the same as the ones considered by us. Furthermore, it is the first time that the performances of approximate methods in reproducing the 3DPS of flux are quantified. 

\cite{Gnedin_Hui_1998} proposed hydro-particle mesh (HPM), a method to describe baryonic pressure as a modification of the gravitational potential in collisionless simulations. They compare the results of their own technique with two reference hydrodynamic simulations. After computing  300 flux skewers at $z=3$, they found out that the mean error on the fractional flux decrement 
is smaller than 10\% in the whole dynamic range. They do not consider other statistics of the flux, but they show that the accuracy in reproducing the column density distribution is around 13\%. They claim that HPM would be suitable when an accuracy of 10-15\% is needed in the modeling. Both our methods and GS+FGPA (with appropriate smoothing length) result
in higher accuracy. 

\cite{Meiksin_2001} also used the HPM technique to compute the flux field. In addition, they considered an $N$-body particle mesh code, from which they computed the flux using GS+FGPA. 
They show that the two methods yield the same cumulative distribution
of the flux within $\sim$10\% accuracy, for four different
cosmological models. The cumulative distributions of column density
and Doppler parameter differ up to $\sim10\%$ and $\sim20\%$, respectively. Although we consider the PDF and not the cumulative PDF of the flux, their findings agree with our results for the PDF given by GS+FGPA.

\cite{Viel_2002} tested GS+FGPA against a
smoothed particle hydrodynamic simulation. Furthermore, they 
developed a
hydro-calibrated approximate method to predict the \lya forest, based
on an adaptive filtering scale for the DM density. Although the
accuracy of these techniques in recovering the logarithm of the flux
PDF given by the hydrodynamic simulation is not quantified, it can be
inferred from their plots that GS+FGPA recovers such
statistics on average better than 10\%, even though the agreement
looks worse in certain regions of flux (e.g. around 0.2 or 0.8). The
PDF  
is reproduced much better by the hydro-calibrated
method. Its accuracy can be estimated through eye-balling to be at percent level. 
It
would then mean that 3D-IMS is comparable to the method provided by
\cite{Viel_2002}, as far as the flux PDF is concerned. 1D-IMS still
performs much better, matching this statistics by construction.

LyMAS method \citep{Peirani_2014} also matches the flux PDF by construction. In its simplest version, this method consists of two hydro-calibrated
transformations of the matter density field. Qualitatively, it can be seen that the method reproduces well the
1DPS given by the reference hydrodynamic simulation, except for $k \gtrsim
1.5\,/\mathrm{cMpc}$. However, LyMAS can be extended with two further
transformations of the flux field (LyMAS full scheme). In this way,
the accuracy of the 1DPS is dramatically improved, although this was
not quantified in \cite{Peirani_2014}. 
Flux skewers appear reasonable only in the full scheme, while in the simplest incarnation they are quite noisy.

Comparing LyMAS to our methods, it certainly does better
than 3D-IMS at reproducing the PDF.
In this respect, it is as good as
1D-IMS, since both match the flux PDF by construction. It also looks like LyMAS recovers the 1DPS given by the hydrodynamic simulation to a very high accuracy. Likewise, 1D-IMS reproduces the 1DPS almost perfectly and accurately reproduces the
3DPS as well. Furthermore, both 3D-IMS and 1D-IMS provide good-looking
skewers applying simple transformations. The LyMAS methodology may be improved by also using the velocity field of the N-body run, which is currently being neglected.

An important feature of 3D-IMS is that it matches the flux provided by the hydrodynamic simulation in real space. This sets the correct filtering scale, allowing us to explore many values of $T_0$ and $\gamma$ when computing the redshift-space flux. On the contrary, LyMAS connects the dark matter density directly with the redshift-space flux, so each choice of $(T_0,\, \gamma, \lambda_G)$ requires an additional hydrodynamic simulation. Therefore, in this regard, 3D-IMS appears to be more flexible than LyMAS.

Recently, \cite{Lochhaas_2015} used LyMAS to predict the cross-correlation between DM halos and \lya forest flux, and compared it to quasars-damped \lya systems cross-correlation measurements from BOSS \citep{Font-Ribera_2012b, Font-Ribera_2013}. From the plots presented, one can tell that the DM halos-\lya forest cross-correlation given by LyMAS reproduces very well the results of their calibrating hydrodynamic simulation (Horizon-AGN; \citealt{Dubois_2014}). However, other statistics relevant for our work, like the 1DPS, are not computed.

\section{Conclusions}
\label{sec:conclusions}

In this work we investigated approximate methods to obtain statistical properties of \lya forest from N-body 
simulations. We focus our attention on the PDF, 1DPS and 3DPS of the flux field,
comparing results of approximate methods with a reference hydrodynamic simulation.

We studied the limitations of the FGPA, which is the basis of many approximate methods.
The primary sources of error are the differences between DM and baryon velocity fields and, to a smaller degree,
the impact of scatter in the temperature-density relationship of the IGM. The accuracy of the FGPA in reproducing the
1DPS and PDF is around 2\%, and around 5\% for the 3DPS.
We also assessed the accuracy of the widely used Gaussian smoothing technique, combined with the FGPA (GS+FGPA).
This method consists in mocking the baryon density by smoothing the matter density with a Gaussian kernel.
Such field is then used to compute the flux within the FGPA. The accuracy at which the statistics of the flux
given by the reference hydrodynamic simulation is reproduced varies a lot with the choice of the smoothing scale $\lambda_G$.
We explored a wide range of smoothing lengths and found out that the best accuracy achieved for 1DPS and 3DPS is
$\sim 7\%$ and $\sim 5\%$, respectively (at $\lambda_G=57\,\mathrm{ckpc}$), and $\sim 4\%$ ($\lambda_G=114 \, \mathrm{ckpc}$)
for the PDF. For smoothing scales $\gtrsim 171\,\mathrm{ckpc}$ the mean accuracy is worse than ~20\%.
This dependence of GS+FGPA on the smoothing scale is rather unfortunate, as the ``optimal''
smoothing scale is guaranteed to differ for models with different thermal IGM history.  As one does not
know a priori this optimal value, in practice it means that works using {\it any} particular smoothing
scale will have error varying in an uncontrolled manner.

To remedy these problems, we have developed two new methods, 3D-IMS and 1D-IMS,
based on the idea of Iteratively Matched Statistics (IMS).
Their starting point is also Gaussian-smoothing the matter density on a certain scale,
which corresponds to the mean interparticle spacing of the simulation considered.
In 3D-IMS, smoothing is followed by matching the 3D
power spectrum and PDF of the flux in real space to the reference hydrodynamic simulation.
With 1D-IMS, we additionally match
the 1DPS and PDF of the flux in redshift space. In contrast to GS+FGPA, 3D-IMS is much less
dependent on the initial smoothing length. It reproduces the 3D power spectrum of the flux in redshift space as
accurate as GS+FGPA when smoothing scales are small, but
performs significantly better for large smoothing scales, with an accuracy of $\sim7\%$
even for smoothing scales as large as  $\sim 500\,\mathrm{ckpc}$. This is a very important property, as it brings
significantly more accurate models of the \lya forest statistics in large-volume simulations where the mean interparticle
spacing has to be large due to computational constrains.
The 1D-IMS method matches flux 1DPS and PDF
by construction. It still performs equally well, or better than GS+FGPA in reproducing the 3DPS ($\sim$5\%).
It is not necessary to use both methods; one can use 3D-IMS only, reproducing the flux 3D power spectrum accurately,
at the expense of a lower accuracy for 1DPS and PDF.

These assessments stand for modeling the \lya forest even in high resolution N-body simulations, but
are especially prominent when large-volume (thus coarse resolution) N-body simulations are used.
We have showed that IMS approximate methods significantly outperform GS+FGPA in such case.
Indeed, through the iterative procedure, our method correctly recovers small-scale physics which is otherwise
not present in low-resolution simulations. In particular, 3D-IMS improves the accuracy in the 1DPS by a factor of 4 with respect
to GS+FGPA. In addition, 3D-IMS appears more robust and easy to implement,
constituting an improvement over previous techniques.

\subsection{Perspectives}
\label{sec:conclusions_persp}

Our methods have applicability in any context where large-box simulations are needed.
The high accuracy of 3D-IMS and 1D-IMS at large smoothing lengths
demonstrates that the hydro-calibrated mappings are able to ``paste'' information about the small-scale
physics of the IGM not present in a large volume simulation, without compromising large-scale statistics.
To be quantitative, at $\lambda_G=228\,\rm ckpc$ 1D-IMS matches perfectly
the flux 1DPS and PDF of the reference hydrodynamic simulation and
recovers the 3DPS within 7\% accuracy. Since in any realistic
situation $\lambda_G$ has to be at least as large as the
inter-particle separation, it means that one would achieve the
aforementioned accuracy applying 1D-IMS to a collisionless simulation
with a trillion particles in a $\sim2.5\,\rm cGpc$ box.
This size is large enough to comfortably study the signature of the BAO signal in the \lya forest.
As a reference, in context of BOSS survey \cite{White_2010} ran a suite of N-body simulations with a box size
of $1.02\,\rm cGpc$ and 4000$^3$ particles (inter-particle separation $260\, \rm ckpc$), applying the Gaussian-smoothing technique.
Currently, the state of the art for N-body simulations is represented by the
``Outer Rim'' (box size $4.3 \, \rm cGpc$, 10240$^3$ particles; \citealt{Habib_2012, Habib_2013}) and
``Dark Sky''  (box size $11.5\, \rm cGpc$, 10240$^3$ particles; \citealt{Skillman_2014}) simulations.

The BAO signal can be modulated by UV background fluctuations, which are coupled to fluctuations in the mean
free path of ionizing photons on large scales \citep{Pontzen_2014, Gontcho-Gontcho_2014}.
For a proper modeling, one needs a simulation with a box size much larger than the mean free path \citep{Davies_2015},
which is of order the BAO scale at $z\sim 2.5$ (\citealt{Worseck_2014} and references therein). Therefore, one would have to
run radiative transfer simulations with box sizes of the order of $1\,\rm cGpc$ --- far beyond current computational
capabilities. The high quasar density in the BOSS survey
allows measuring the 3D power spectrum, which can be exploited to improve cosmological constrains and/or
constrain IGM thermal properties \citep{McQuinn_2011, McQuinn_2011b}.
Finally, our technique could help in modeling the cross-correlation between
\lya forest and ${\rm HI}$ $21\, \rm cm$ signal \citep{Guha_2015}, as well as between CMB lensing and \lya forest
\citep{Vallinotto_2009, Vallinotto_2011}.

\acknowledgments 

We thank Martin White, Casey W. Stark and the members of the ENIGMA group at the
Max Planck Institute for Astronomy (MPIA) for helpful comments and discussions.
We are grateful to Vetter's Alt Heidelberger for providing supportive and enriching 
environment for many of those discussions. We thank the Esalen Institute for the kind hospitality. 
Calculations presented in this paper used resources of the National Energy Research Scientific Computing Center (NERSC),
which is supported by the Office of Science of the U.S. Department of Energy under Contract No. DE-AC02-05CH11231.
Hydrodynamical runs in this paper were done under the ASCR Leadership Computing Challenge (ALCC) allocation.
ZL acknowledges support from the Scientific Discovery through Advanced
Computing (SciDAC) program funded by U.S.\ Department of Energy Office of
Advanced Scientific Computing Research and the Office of High Energy Physics.
This work made extensive use of the NASA Astrophysics Data System and of the astro-ph preprint archive at arXiv.org.

\bibliographystyle{apj}
\bibliography{Sorini_25Feb2016}

\begin{thebibliography}{64}
\expandafter\ifx\csname natexlab\endcsname\relax\def\natexlab#1{#1}\fi

\bibitem[{{Almgren} {et~al.}(2013){Almgren}, {Bell}, {Lijewski}, {Luki{\'c}},
  \& {Van Andel}}]{Almgren_2013}
{Almgren}, A.~S., {Bell}, J.~B., {Lijewski}, M.~J., {Luki{\'c}}, Z., \& {Van
  Andel}, E. 2013, \apj, 765, 39

\bibitem[{{Aragon-Calvo} \& {Szalay}(2013)}]{Argon-Calvo_2013}
{Aragon-Calvo}, M.~A., \& {Szalay}, A.~S. 2013, \mnras, 428, 3409

\bibitem[{{Barnes} \& {Hut}(1986)}]{Barnes_1986}
{Barnes}, J., \& {Hut}, P. 1986, \nat, 324, 446

\bibitem[{{Becker} \& {Bolton}(2013)}]{Becker_2013}
{Becker}, G.~D., \& {Bolton}, J.~S. 2013, \mnras, 436, 1023

\bibitem[{{Binney} \& {Tremaine}(2008)}]{Binney_2008}
{Binney}, J., \& {Tremaine}, S. 2008, {Galactic Dynamics: Second Edition}
  (Princeton University Press)

\bibitem[{{Brown} {et~al.}(1989){Brown}, {Byrne}, \& {Hindmarsh}}]{Brown_1989}
{Brown}, P.~N., {Byrne}, G.~D., \& {Hindmarsh}, A.~C. 1989, SIAM J. Sci. Stat.
  Comput., 10, 1038

\bibitem[{{Cen} {et~al.}(1994){Cen}, {Miralda-Escud{\'e}}, {Ostriker}, \&
  {Rauch}}]{Cen_1994}
{Cen}, R., {Miralda-Escud{\'e}}, J., {Ostriker}, J.~P., \& {Rauch}, M. 1994,
  \apjl, 437, L9

\bibitem[{{Colella}(1990)}]{Colella_1990}
{Colella}, P. 1990, Journal of Computational Physics, 87, 171

\bibitem[{{Croft} {et~al.}(2002){Croft}, {Hernquist}, {Springel}, {Westover},
  \& {White}}]{Croft_2002}
{Croft}, R.~A.~C., {Hernquist}, L., {Springel}, V., {Westover}, M., \& {White},
  M. 2002, \apj, 580, 634

\bibitem[{{Croft} {et~al.}(1998){Croft}, {Weinberg}, {Katz}, \&
  {Hernquist}}]{Croft_1998}
{Croft}, R.~A.~C., {Weinberg}, D.~H., {Katz}, N., \& {Hernquist}, L. 1998, in
  Large Scale Structure: Tracks and Traces, ed. V.~{Mueller}, S.~{Gottloeber},
  J.~P. {Muecket}, \& J.~{Wambsganss}, 69--75

\bibitem[{{Croft} {et~al.}(1999){Croft}, {Weinberg}, {Pettini}, {Hernquist}, \&
  {Katz}}]{Croft_1999}
{Croft}, R.~A.~C., {Weinberg}, D.~H., {Pettini}, M., {Hernquist}, L., \&
  {Katz}, N. 1999, \apj, 520, 1

\bibitem[{{Davies} \& {Furlanetto}(2015)}]{Davies_2015}
{Davies}, F.~B., \& {Furlanetto}, S.~R. 2015, ArXiv e-prints

\bibitem[{{Dubois} {et~al.}(2014){Dubois}, {Pichon}, {Welker}, {Le Borgne},
  {Devriendt}, {Laigle}, {Codis}, {Pogosyan}, {Arnouts}, {Benabed}, {Bertin},
  {Blaizot}, {Bouchet}, {Cardoso}, {Colombi}, {de Lapparent}, {Desjacques},
  {Gavazzi}, {Kassin}, {Kimm}, {McCracken}, {Milliard}, {Peirani}, {Prunet},
  {Rouberol}, {Silk}, {Slyz}, {Sousbie}, {Teyssier}, {Tresse}, {Treyer},
  {Vibert}, \& {Volonteri}}]{Dubois_2014}
{Dubois}, Y., {Pichon}, C., {Welker}, C., {Le Borgne}, D., {Devriendt}, J.,
  {Laigle}, C., {Codis}, S., {Pogosyan}, D., {Arnouts}, S., {Benabed}, K.,
  {Bertin}, E., {Blaizot}, J., {Bouchet}, F., {Cardoso}, J.-F., {Colombi}, S.,
  {de Lapparent}, V., {Desjacques}, V., {Gavazzi}, R., {Kassin}, S., {Kimm},
  T., {McCracken}, H., {Milliard}, B., {Peirani}, S., {Prunet}, S., {Rouberol},
  S., {Silk}, J., {Slyz}, A., {Sousbie}, T., {Teyssier}, R., {Tresse}, L.,
  {Treyer}, M., {Vibert}, D., \& {Volonteri}, M. 2014, \mnras, 444, 1453

\bibitem[{{Faucher-Gigu{\`e}re} {et~al.}(2008){Faucher-Gigu{\`e}re}, {Lidz},
  {Hernquist}, \& {Zaldarriaga}}]{Faucher-Giguere_2008}
{Faucher-Gigu{\`e}re}, C.-A., {Lidz}, A., {Hernquist}, L., \& {Zaldarriaga}, M.
  2008, \apj, 688, 85

\bibitem[{{Font-Ribera} {et~al.}(2013){Font-Ribera}, {Arnau},
  {Miralda-Escud{\'e}}, {Rollinde}, {Brinkmann}, {Brownstein}, {Lee}, {Myers},
  {Palanque-Delabrouille}, {P{\^a}ris}, {Petitjean}, {Rich}, {Ross},
  {Schneider}, \& {White}}]{Font-Ribera_2013}
{Font-Ribera}, A., {Arnau}, E., {Miralda-Escud{\'e}}, J., {Rollinde}, E.,
  {Brinkmann}, J., {Brownstein}, J.~R., {Lee}, K.-G., {Myers}, A.~D.,
  {Palanque-Delabrouille}, N., {P{\^a}ris}, I., {Petitjean}, P., {Rich}, J.,
  {Ross}, N.~P., {Schneider}, D.~P., \& {White}, M. 2013, \jcap, 5, 18

\bibitem[{{Font-Ribera} {et~al.}(2014){Font-Ribera}, {McDonald}, {Mostek},
  {Reid}, {Seo}, \& {Slosar}}]{Font-Ribera_2014}
{Font-Ribera}, A., {McDonald}, P., {Mostek}, N., {Reid}, B.~A., {Seo}, H.-J.,
  \& {Slosar}, A. 2014, \jcap, 5, 023

\bibitem[{{Font-Ribera} {et~al.}(2012){Font-Ribera}, {Miralda-Escud{\'e}},
  {Arnau}, {Carithers}, {Lee}, {Noterdaeme}, {P{\^a}ris}, {Petitjean}, {Rich},
  {Rollinde}, {Ross}, {Schneider}, {White}, \& {York}}]{Font-Ribera_2012b}
{Font-Ribera}, A., {Miralda-Escud{\'e}}, J., {Arnau}, E., {Carithers}, B.,
  {Lee}, K.-G., {Noterdaeme}, P., {P{\^a}ris}, I., {Petitjean}, P., {Rich}, J.,
  {Rollinde}, E., {Ross}, N.~P., {Schneider}, D.~P., {White}, M., \& {York},
  D.~G. 2012, \jcap, 11, 59

\bibitem[{{Furlanetto} {et~al.}(2006){Furlanetto}, {Oh}, \&
  {Briggs}}]{Furlanetto_2006}
{Furlanetto}, S.~R., {Oh}, S.~P., \& {Briggs}, F.~H. 2006, \physrep, 433, 181

\bibitem[{{Gnedin} \& {Hui}(1996)}]{Gnedin_1996}
{Gnedin}, N.~Y., \& {Hui}, L. 1996, \apjl, 472, L73

\bibitem[{{Gnedin} \& {Hui}(1998)}]{Gnedin_Hui_1998}
---. 1998, \mnras, 296, 44

\bibitem[{{Gontcho A Gontcho} {et~al.}(2014){Gontcho A Gontcho},
  {Miralda-Escud{\'e}}, \& {Busca}}]{Gontcho-Gontcho_2014}
{Gontcho A Gontcho}, S., {Miralda-Escud{\'e}}, J., \& {Busca}, N.~G. 2014,
  \mnras, 442, 187

\bibitem[{{Guha Sarkar} \& {Datta}(2015)}]{Guha_2015}
{Guha Sarkar}, T., \& {Datta}, K.~K. 2015, \jcap, 8, 1

\bibitem[{{Haardt} \& {Madau}(2012)}]{Haardt_Madau_2012}
{Haardt}, F., \& {Madau}, P. 2012, \apj, 746, 125

\bibitem[{{Habib} {et~al.}(2012){Habib}, {Morozov}, {Finkel}, {Pope},
  {Heitmann}, {Kumaran}, {Peterka}, {Insley}, {Daniel}, {Fasel}, {Frontiere},
  \& {Lukic}}]{Habib_2012}
{Habib}, S., {Morozov}, V., {Finkel}, H., {Pope}, A., {Heitmann}, K.,
  {Kumaran}, K., {Peterka}, T., {Insley}, J., {Daniel}, D., {Fasel}, P.,
  {Frontiere}, N., \& {Lukic}, Z. 2012, ArXiv e-prints

\bibitem[{{Habib} {et~al.}(2013){Habib}, {Morozov}, {Frontiere}, {Finkel},
  {Pope}, \& {Heitmann}}]{Habib_2013}
{Habib}, S., {Morozov}, V., {Frontiere}, N., {Finkel}, H., {Pope}, A., \&
  {Heitmann}, K. 2013, in SC '13 Proceedings of SC13: International Conference
  for High Performance Computing, Networking, Storage and Analysis, Article \#6

\bibitem[{{Hernquist} {et~al.}(1996){Hernquist}, {Katz}, {Weinberg}, \&
  {Miralda-Escud{\'e}}}]{Hernquist_1996}
{Hernquist}, L., {Katz}, N., {Weinberg}, D.~H., \& {Miralda-Escud{\'e}}, J.
  1996, \apjl, 457, L51

\bibitem[{{Hui} \& {Gnedin}(1997)}]{Hui_1997}
{Hui}, L., \& {Gnedin}, N.~Y. 1997, \mnras, 292, 27

\bibitem[{{Kaurov} \& {Gnedin}(2014)}]{Kaurov_2014}
{Kaurov}, A.~A., \& {Gnedin}, N.~Y. 2014, \apj, 787, 146

\bibitem[{{Komatsu} {et~al.}(2011){Komatsu}, {Smith}, {Dunkley}, {Bennett},
  {Gold}, {Hinshaw}, {Jarosik}, {Larson}, {Nolta}, {Page}, {Spergel},
  {Halpern}, {Hill}, {Kogut}, {Limon}, {Meyer}, {Odegard}, {Tucker}, {Weiland},
  {Wollack}, \& {Wright}}]{WMAP}
{Komatsu}, E., {Smith}, K.~M., {Dunkley}, J., {Bennett}, C.~L., {Gold}, B.,
  {Hinshaw}, G., {Jarosik}, N., {Larson}, D., {Nolta}, M.~R., {Page}, L.,
  {Spergel}, D.~N., {Halpern}, M., {Hill}, R.~S., {Kogut}, A., {Limon}, M.,
  {Meyer}, S.~S., {Odegard}, N., {Tucker}, G.~S., {Weiland}, J.~L., {Wollack},
  E., \& {Wright}, E.~L. 2011, \apjs, 192, 18

\bibitem[{{Kuhlen} \& {Faucher-Gigu{\`e}re}(2012)}]{Kuhlen_2012}
{Kuhlen}, M., \& {Faucher-Gigu{\`e}re}, C.-A. 2012, \mnras, 423, 862

\bibitem[{{Kulkarni} {et~al.}(2015){Kulkarni}, {Hennawi}, {O{\~n}orbe},
  {Rorai}, \& {Springel}}]{Kulkarni_2015}
{Kulkarni}, G., {Hennawi}, J.~F., {O{\~n}orbe}, J., {Rorai}, A., \& {Springel},
  V. 2015, ArXiv e-prints

\bibitem[{{Lee}(2012)}]{Lee_2012}
{Lee}, K.-G. 2012, \apj, 753, 136

\bibitem[{{Lee} {et~al.}(2015){Lee}, {Hennawi}, {Spergel}, {Weinberg}, {Hogg},
  {Viel}, {Bolton}, {Bailey}, {Pieri}, {Carithers}, {Schlegel}, {Lundgren},
  {Palanque-Delabrouille}, {Suzuki}, {Schneider}, \& {Y{\`e}che}}]{Lee_2015}
{Lee}, K.-G., {Hennawi}, J.~F., {Spergel}, D.~N., {Weinberg}, D.~H., {Hogg},
  D.~W., {Viel}, M., {Bolton}, J.~S., {Bailey}, S., {Pieri}, M.~M.,
  {Carithers}, W., {Schlegel}, D.~J., {Lundgren}, B., {Palanque-Delabrouille},
  N., {Suzuki}, N., {Schneider}, D.~P., \& {Y{\`e}che}, C. 2015, \apj, 799, 196

\bibitem[{{Lidz} {et~al.}(2010){Lidz}, {Faucher-Gigu{\`e}re}, {Dall'Aglio},
  {McQuinn}, {Fechner}, {Zaldarriaga}, {Hernquist}, \& {Dutta}}]{Lidz_2010}
{Lidz}, A., {Faucher-Gigu{\`e}re}, C.-A., {Dall'Aglio}, A., {McQuinn}, M.,
  {Fechner}, C., {Zaldarriaga}, M., {Hernquist}, L., \& {Dutta}, S. 2010, \apj,
  718, 199

\bibitem[{{Lochhaas} {et~al.}(2015){Lochhaas}, {Weinberg}, {Peirani}, {Dubois},
  {Colombi}, {Blaizot}, {Font-Ribera}, {Pichon}, \&
  {Devriendt}}]{Lochhaas_2015}
{Lochhaas}, C., {Weinberg}, D.~H., {Peirani}, S., {Dubois}, Y., {Colombi}, S.,
  {Blaizot}, J., {Font-Ribera}, A., {Pichon}, C., \& {Devriendt}, J. 2015,
  ArXiv e-prints

\bibitem[{{Luki{\'c}} {et~al.}(2015){Luki{\'c}}, {Stark}, {Nugent}, {White},
  {Meiksin}, \& {Almgren}}]{Lukic_2015}
{Luki{\'c}}, Z., {Stark}, C.~W., {Nugent}, P., {White}, M., {Meiksin}, A.~A.,
  \& {Almgren}, A. 2015, \mnras, 446, 3697

\bibitem[{{McDonald} {et~al.}(2000){McDonald}, {Miralda-Escud{\'e}}, {Rauch},
  {Sargent}, {Barlow}, {Cen}, \& {Ostriker}}]{McDonald_2000}
{McDonald}, P., {Miralda-Escud{\'e}}, J., {Rauch}, M., {Sargent}, W.~L.~W.,
  {Barlow}, T.~A., {Cen}, R., \& {Ostriker}, J.~P. 2000, \apj, 543, 1

\bibitem[{{McQuinn} {et~al.}(2011){McQuinn}, {Hernquist}, {Lidz}, \&
  {Zaldarriaga}}]{McQuinn_2011}
{McQuinn}, M., {Hernquist}, L., {Lidz}, A., \& {Zaldarriaga}, M. 2011, \mnras,
  415, 977

\bibitem[{{McQuinn} \& {White}(2011)}]{McQuinn_2011b}
{McQuinn}, M., \& {White}, M. 2011, \mnras, 415, 2257

\bibitem[{{Meiksin} \& {White}(2001)}]{Meiksin_2001}
{Meiksin}, A., \& {White}, M. 2001, \mnras, 324, 141

\bibitem[{{Meiksin}(2009)}]{Meiksin_review}
{Meiksin}, A.~A. 2009, Reviews of Modern Physics, 81, 1405

\bibitem[{{Miralda-Escud{\'e}}(2003)}]{Miralda_2003}
{Miralda-Escud{\'e}}, J. 2003, \apj, 597, 66

\bibitem[{{Norman} {et~al.}(2009){Norman}, {Paschos}, \&
  {Harkness}}]{Norman_2009}
{Norman}, M.~L., {Paschos}, P., \& {Harkness}, R. 2009, Journal of Physics
  Conference Series, 180, 012021

\bibitem[{{Peirani} {et~al.}(2014){Peirani}, {Weinberg}, {Colombi}, {Blaizot},
  {Dubois}, \& {Pichon}}]{Peirani_2014}
{Peirani}, S., {Weinberg}, D.~H., {Colombi}, S., {Blaizot}, J., {Dubois}, Y.,
  \& {Pichon}, C. 2014, \apj, 784, 11

\bibitem[{{Petitjean} {et~al.}(1995){Petitjean}, {Mueket}, \&
  {Kates}}]{Petitjean_1995}
{Petitjean}, P., {Mueket}, J.~P., \& {Kates}, R.~E. 1995, \aap, 295, L9

\bibitem[{{Pontzen}(2014)}]{Pontzen_2014}
{Pontzen}, A. 2014, \prd, 89, 083010

\bibitem[{{Rauch}(1998)}]{Rauch_1998}
{Rauch}, M. 1998, \araa, 36, 267

\bibitem[{{Rauch} {et~al.}(1997){Rauch}, {Miralda-Escud{\'e}}, {Sargent},
  {Barlow}, {Weinberg}, {Hernquist}, {Katz}, {Cen}, \& {Ostriker}}]{Rauch_1997}
{Rauch}, M., {Miralda-Escud{\'e}}, J., {Sargent}, W.~L.~W., {Barlow}, T.~A.,
  {Weinberg}, D.~H., {Hernquist}, L., {Katz}, N., {Cen}, R., \& {Ostriker},
  J.~P. 1997, \apj, 489, 7

\bibitem[{{Rorai} {et~al.}(2013){Rorai}, {Hennawi}, \& {White}}]{Rorai_2013}
{Rorai}, A., {Hennawi}, J.~F., \& {White}, M. 2013, \apj, 775, 81

\bibitem[{{Skillman} {et~al.}(2014){Skillman}, {Warren}, {Turk}, {Wechsler},
  {Holz}, \& {Sutter}}]{Skillman_2014}
{Skillman}, S.~W., {Warren}, M.~S., {Turk}, M.~J., {Wechsler}, R.~H., {Holz},
  D.~E., \& {Sutter}, P.~M. 2014, ArXiv e-prints

\bibitem[{{Slosar} {et~al.}(2011){Slosar}, {Font-Ribera}, {Pieri}, {Rich}, {Le
  Goff}, {Aubourg}, {Brinkmann}, {Busca}, {Carithers}, {Charlassier},
  {Cort{\^e}s}, {Croft}, {Dawson}, {Eisenstein}, {Hamilton}, {Ho}, {Lee},
  {Lupton}, {McDonald}, {Medolin}, {Muna}, {Miralda-Escud{\'e}}, {Myers},
  {Nichol}, {Palanque-Delabrouille}, {P{\^a}ris}, {Petitjean}, {Pi{\v s}kur},
  {Rollinde}, {Ross}, {Schlegel}, {Schneider}, {Sheldon}, {Weaver}, {Weinberg},
  {Yeche}, \& {York}}]{Slosar_2011}
{Slosar}, A., {Font-Ribera}, A., {Pieri}, M.~M., {Rich}, J., {Le Goff}, J.-M.,
  {Aubourg}, {\'E}., {Brinkmann}, J., {Busca}, N., {Carithers}, B.,
  {Charlassier}, R., {Cort{\^e}s}, M., {Croft}, R., {Dawson}, K.~S.,
  {Eisenstein}, D., {Hamilton}, J.-C., {Ho}, S., {Lee}, K.-G., {Lupton}, R.,
  {McDonald}, P., {Medolin}, B., {Muna}, D., {Miralda-Escud{\'e}}, J., {Myers},
  A.~D., {Nichol}, R.~C., {Palanque-Delabrouille}, N., {P{\^a}ris}, I.,
  {Petitjean}, P., {Pi{\v s}kur}, Y., {Rollinde}, E., {Ross}, N.~P.,
  {Schlegel}, D.~J., {Schneider}, D.~P., {Sheldon}, E., {Weaver}, B.~A.,
  {Weinberg}, D.~H., {Yeche}, C., \& {York}, D.~G. 2011, \jcap, 9, 001

\bibitem[{{Slosar} {et~al.}(2009){Slosar}, {Ho}, {White}, \&
  {Louis}}]{Slosar_2009}
{Slosar}, A., {Ho}, S., {White}, M., \& {Louis}, T. 2009, \jcap, 10, 19

\bibitem[{{Springel}(2005)}]{Springel_2005}
{Springel}, V. 2005, \mnras, 364, 1105

\bibitem[{{Strang}(1968)}]{Strang_1968}
{Strang}, G. 1968, SIAM Journal on Numerical Analysis, 5, 506

\bibitem[{{Vallinotto} {et~al.}(2009){Vallinotto}, {Das}, {Spergel}, \&
  {Viel}}]{Vallinotto_2009}
{Vallinotto}, A., {Das}, S., {Spergel}, D.~N., \& {Viel}, M. 2009, Physical
  Review Letters, 103, 091304

\bibitem[{{Vallinotto} {et~al.}(2011){Vallinotto}, {Viel}, {Das}, \&
  {Spergel}}]{Vallinotto_2011}
{Vallinotto}, A., {Viel}, M., {Das}, S., \& {Spergel}, D.~N. 2011, \apj, 735,
  38

\bibitem[{{van de Weygaert} \& {van Kampen}(1993)}]{van_de_Weygaert_1993}
{van de Weygaert}, R., \& {van Kampen}, E. 1993, \mnras, 263, 481

\bibitem[{{Viel} {et~al.}(2006){Viel}, {Haehnelt}, \& {Springel}}]{Viel_2006}
{Viel}, M., {Haehnelt}, M.~G., \& {Springel}, V. 2006, \mnras, 367, 1655

\bibitem[{{Viel} {et~al.}(2002){Viel}, {Matarrese}, {Mo}, {Theuns}, \&
  {Haehnelt}}]{Viel_2002}
{Viel}, M., {Matarrese}, S., {Mo}, H.~J., {Theuns}, T., \& {Haehnelt}, M.~G.
  2002, \mnras, 336, 685

\bibitem[{{Weinberg} {et~al.}(1997){Weinberg}, {Hernsquit}, {Katz}, {Croft}, \&
  {Miralda-Escud{\'e}}}]{Weinberg_1997}
{Weinberg}, D.~H., {Hernsquit}, L., {Katz}, N., {Croft}, R., \&
  {Miralda-Escud{\'e}}, J. 1997, in Structure and Evolution of the
  Intergalactic Medium from QSO Absorption Line System, ed. P.~{Petitjean} \&
  S.~{Charlot}, 133

\bibitem[{{White} {et~al.}(2010){White}, {Pope}, {Carlson}, {Heitmann},
  {Habib}, {Fasel}, {Daniel}, \& {Lukic}}]{White_2010}
{White}, M., {Pope}, A., {Carlson}, J., {Heitmann}, K., {Habib}, S., {Fasel},
  P., {Daniel}, D., \& {Lukic}, Z. 2010, \apj, 713, 383

\bibitem[{{Worseck} {et~al.}(2014){Worseck}, {Prochaska}, {O'Meara}, {Becker},
  {Ellison}, {Lopez}, {Meiksin}, {M{\'e}nard}, {Murphy}, \&
  {Fumagalli}}]{Worseck_2014}
{Worseck}, G., {Prochaska}, J.~X., {O'Meara}, J.~M., {Becker}, G.~D.,
  {Ellison}, S.~L., {Lopez}, S., {Meiksin}, A., {M{\'e}nard}, B., {Murphy},
  M.~T., \& {Fumagalli}, M. 2014, \mnras, 445, 1745

\bibitem[{{Zel'dovich}(1970)}]{Zeldovich_1970}
{Zel'dovich}, Y.~B. 1970, \aap, 5, 84

\bibitem[{{Zhang} {et~al.}(1997){Zhang}, {Anninos}, {Norman}, \&
  {Meiksin}}]{Zhang_1997}
{Zhang}, Y., {Anninos}, P., {Norman}, M.~L., \& {Meiksin}, A. 1997, \apj, 485,
  496

\end{thebibliography}

\appendix
\section{Optimal Smoothing Length for Velocity}
\label{app:smoothing_vel}

The baryon density field can be mocked through a Gaussian smoothing of the DM density field. Also the velocity field of DM should be smoothed accordingly to reproduce the velocity field of baryons. In principle, there might be two different optimal values of $\lambda_G$ for density and velocity, so one should vary $\lambda_G$ for both quantities and explore all possible combinations within the dynamic range considered. However, this extensive study can be quite time consuming and is probably not the most efficient way to proceed. In all techniques tested, we shall vary $\lambda_G$ for the density, keeping it fixed for the velocity. We determine the optimal fixed smoothing length for the velocity field as follows. We apply the FGPA (equations \eqref{eq:n_HI} and\eqref{eq:tau}) using the baryon density fluctuations given by our reference hydrodynamic simulation, but smoothing the DM velocity field from the same simulation at different values of $\lambda_G$. We then choose the smoothing length best matching the flux 1DPS, PDF and 3DPS given by the hydrodynamic simulation. 

\begin{figure*}
\centering
\includegraphics[width=\textwidth]{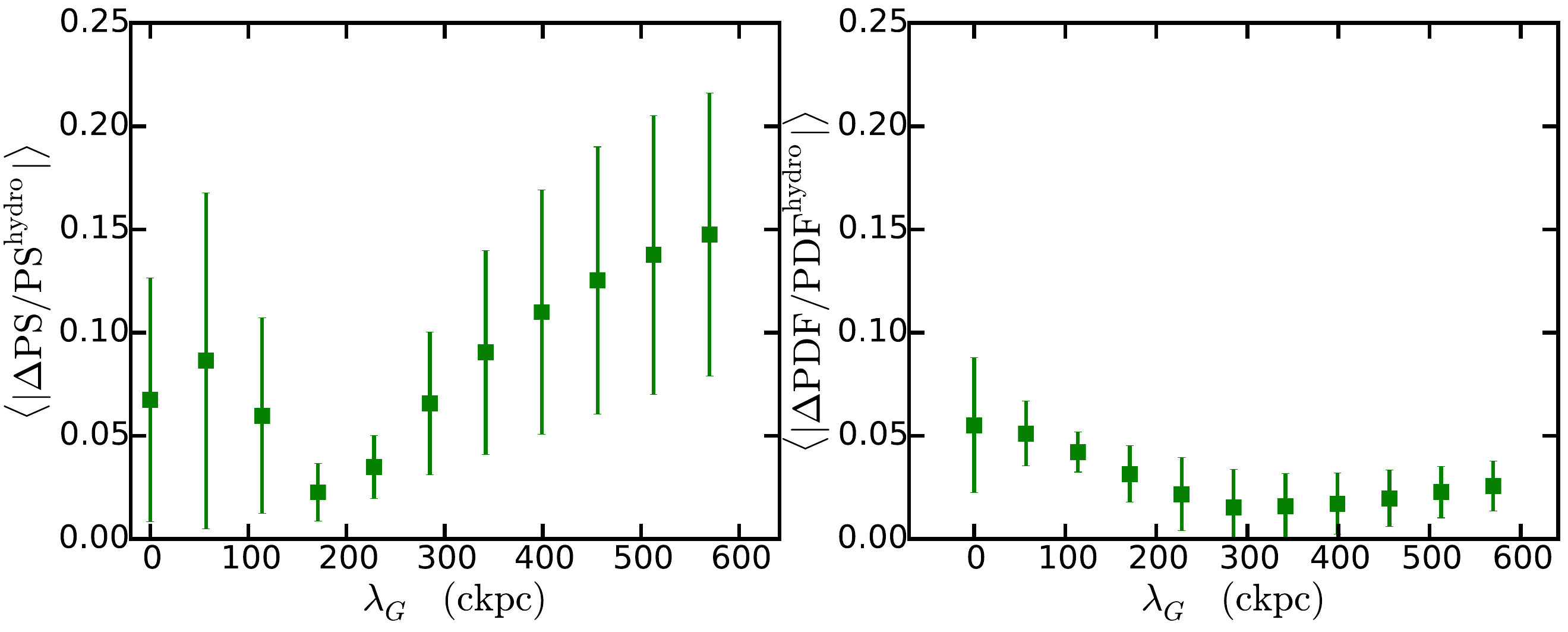}
\caption{Left panel shows the relative error between the dimensionless line-of-sight power spectrum of the flux given by the reference hydrodynamic simulation and of the flux computed by applying a 1-to-1 temperature-density relationship to the baryon density field and using the Gaussian-smoothed line-of-sight velocities of dark matter instead of baryons, as a function of different smoothing lengths of the DM velocity field. Squares mark the mean values of the accuracy, while error bars represent the root-mean-square of the accuracy in the dynamic ranges considered. The plot in the right panel is analogous, but it refers to the accuracy in reproducing the PDF of the flux in redshift space.} 
\label{fig:optimal_velocity}
\centering
\includegraphics[width=\textwidth]{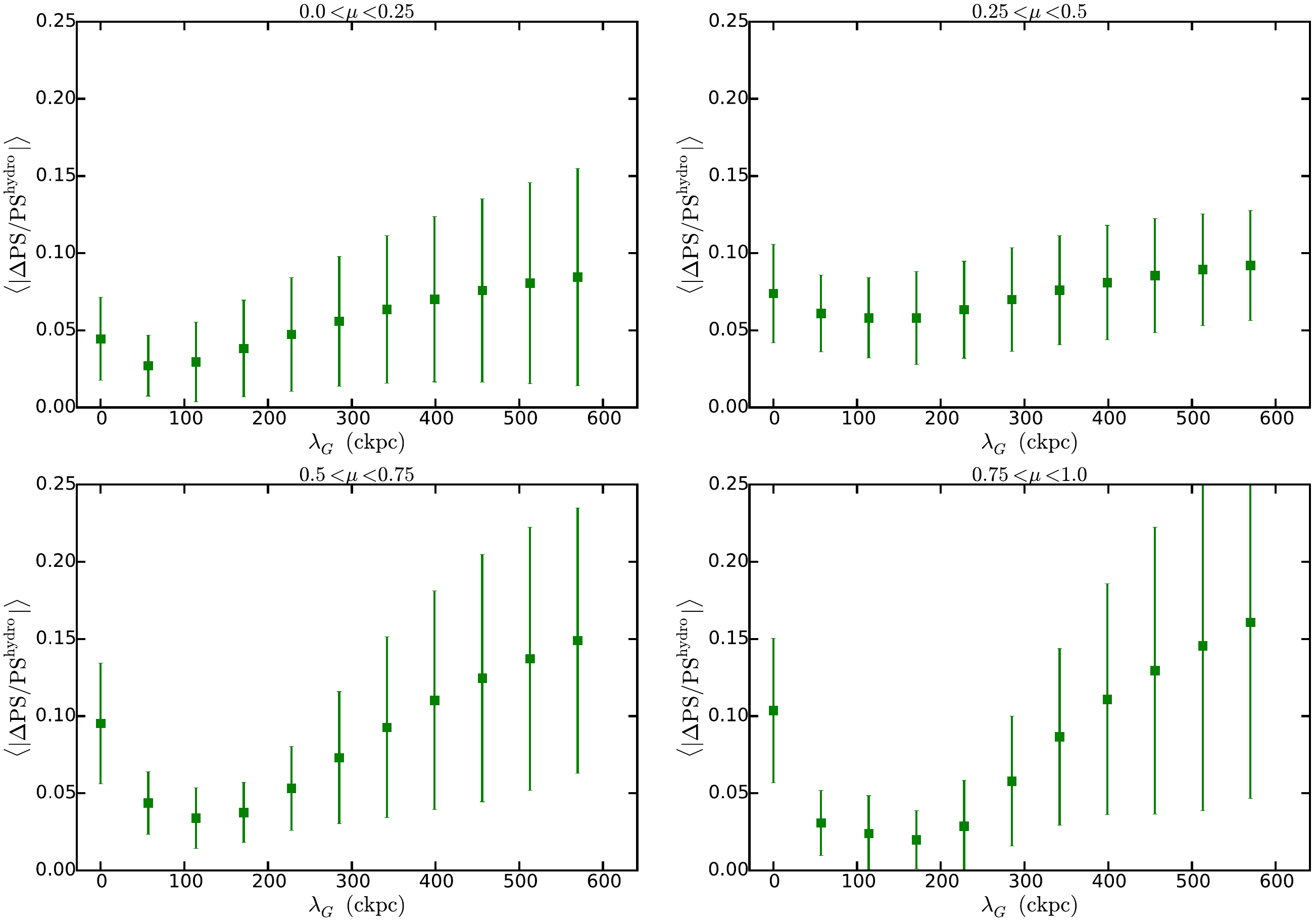}
\caption{Relative error between the dimensionless 3D power spectrum $\Delta^2(k,\,\mu)$ of the flux given by the reference hydrodynamic simulation and of the flux computed by applying a 1-to-1 temperature-density relationship to the baryon density field and using the Gaussian-smoothed line-of-sight velocities of dark matter instead of baryons, as a function of different smoothing lengths of the DM velocity field. Squares mark the mean values of the accuracy, while error bars represent the root-mean-square of the accuracy in the dynamic ranges considered. Each panel refers to a different $\mu$-bin.} 
\label{fig:optimal_velocity_3DPS}
\end{figure*}

The outcome of our analysis can be seen in Figure \ref{fig:optimal_velocity}. In the left panel, we plot the accuracy in reproducing the 1DPS of the hydrodynamic simulation, versus the smoothing length. The right panel displays the analogous plot for the PDF. We notice that the smoothing of the velocities has a strong impact on the 1DPS, the optimal value being $171\,\mathrm{ckpc}$, for which the mean accuracy is $\sim$1\%. The PDF is less sensitive to the smoothing length for $\lambda_G \gtrsim 285\,\mathrm{ckpc}$. 

In Figure \ref{fig:optimal_velocity_3DPS} we show the results for the 3DPS. We see that the impact of the smoothing length is more important for larger $\mu$, whereas $\lambda_G \lesssim 228\,\mathrm{ckpc}$ yield a better accuracy with respect to $\lambda_G \gtrsim 228 \mathrm{ckpc}$. Given the different trends of the 1DPS, 3DPS and PDF accuracy, there is no unique optimal value of $\lambda_G$ to maximize the accuracy in all statistics, so we have to make a compromise. We chose $\lambda_G=228\,\mathrm{ckpc}/h$, for which the accuracy of both PS and PDF is $\sim$2\% and the accuracy of the 3DPS is between 3\% and 5\%, depending on the $\mu$-bin considered. We kept this value fixed in all our work.

Our choice of optimizing the smoothing length of the velocity allows us to focus our analysis on the impact of the smoothing length of the density field on the accuracy of our methods (see section \ref{sec:comparison} for details). As a consequence, the errors quoted for the techniques considered are minimized. Indeed, when we show the error for $\lambda_G<228\,\mathrm{ckpc}$ in the density field we are still smoothing the velocity at $228\,\mathrm{ckpc}$. We recall that the smoothing length has to be at least as large as the inter-particle separation of the simulation adopted. If such separation is $228\,\mathrm{ckpc}$, one can smooth the velocity field at this value and still adopted a larger $\lambda_G$ for the DM density. Conversely, if the inter-particle separation is smaller than $228\,\mathrm{ckpc}$, one can smooth the DM density choosing $\lambda_G$ to be equal to such separation. One can still smooth the velocity field at $228\,\mathrm{ckpc}$, without losing much information. Indeed, the velocity field in voids (which are the most important regions in terms of the flux statistics due to the exponentiation of equation \eqref{eq:tau_real}) is very smooth for very different cosmological models \citep{van_de_Weygaert_1993, Argon-Calvo_2013}. As such, there is no conspicuous small-scale structure in the velocity field (see also Figure \ref{fig:skewers_FGPA}), which is thus less affected by the smoothing than the DM density field. So, it is sensible to consider a fixed smoothing length for the velocity field but varying it for the DM density. 

To get a sense of the accuracy obtained with a certain method adopting a different smoothing length for the velocity field, one can sum in quadrature of the errors reported in Figure \ref{fig:accuracy_1D} and \ref{fig:accuracy_3DPS} with the errors shown in the corresponding plots in this section, i.e. Figure \ref{fig:optimal_velocity} and \ref{fig:optimal_velocity_3DPS}. Doing so, the mean accuracy of our methods would decrease, but the trend of the accuracy versus the smoothing length would be unaffected. In particular, except for the bin along the line-of-sight of the 3DPS, GS+FGPA would still look worse than our methods.  

\end{document}